\documentclass[a4paper,12pt]{article}

\usepackage[dvipdfmx]{graphicx}
\usepackage{amsmath}
\usepackage{amssymb}
\usepackage{amsfonts}
\usepackage{amsthm}
\usepackage{cite}
\usepackage{extarrows}
\usepackage{color}
\usepackage{ulem}
\theoremstyle{definition}

\newtheorem{axiomC}{Axiom} 
\newtheorem{axiomV}{Axiom} 

\newcommand{\dop}{\mathrm{d}} 
\newcommand{\iop}{\iota} 
\newcommand{\pbra}[1]{\boldsymbol{(}#1\boldsymbol{)}_+}
\newcommand{\mbra}[1]{\boldsymbol{(}#1\boldsymbol{)}_-}
\newcommand{\pmbra}[1]{\boldsymbol{(}#1\boldsymbol{)}_\pm}
\newcommand{\bbra}[1]{\boldsymbol{(}#1\boldsymbol{)}}

\date{empty}
\begin{document}
\begin{titlepage}
\null
\begin{flushright}
August, 2020
\end{flushright}
\vskip 2cm
\begin{center}
  {\Large \bf
More on Doubled Aspects of Algebroids \\
\vspace{0.5cm}
in Double Field Theory
}
\vskip 1.5cm
\normalsize
\renewcommand\thefootnote{\alph{footnote}}

{\large
Haruka Mori\footnote{h.mori(at)sci.kitasato-u.ac.jp}
and
Shin Sasaki\footnote{shin-s(at)kitasato-u.ac.jp}
}
\vskip 0.7cm
  {\it
  Department of Physics,  Kitasato University \\
  Sagamihara 252-0373, Japan
}
\vskip 2cm
\begin{abstract}
We continue to study doubled aspects of algebroid structures equipped
 with the {\sf C}-bracket in double field theory (DFT).
We find that a family of algebroids, the Vaisman (metric or pre-DFT), the pre- and the ante-Courant
algebroids are constructed by the analogue of the Drinfel'd double of Lie algebroid pairs.
We examine geometric implementations of these algebroids in the para-Hermitian
 manifold, which is a realization of the doubled space-time in DFT.
We show that the strong constraint in DFT is necessary to realize the
 doubled and non-trivial Poisson structures but can be relaxed for some algebroids.
The doubled structures of twisted brackets and those associated with
 group manifolds are briefly discussed.

\end{abstract}
\end{center}

\end{titlepage}

\newpage
\setcounter{footnote}{0}
\renewcommand\thefootnote{\arabic{footnote}}
\pagenumbering{arabic}
\section{Introduction} \label{sect:introduction}
String theory is a candidate of quantum gravity theory.
Since the fundamental string probes space-time as a one-dimensional object,
the geometry in string theory exhibits quite
different properties compared with that of Einstein gravity.
Dualities, that relate various consistent superstring theories, play
important roles to explore the geometry behind string theories.
Among other things, T-duality is one of the most familiar duality in string
theory. When a closed string winds around compact space directions, the
energy spectrum depends on the discretized Kaluza-Klein (KK) momentum
modes $n \in \mathbb{Z}$ and the winding modes $w \in \mathbb{Z}$ of the
string.
The energy density is invariant under the exchange of $n$ and $w$
together with the inversion of the radius $R$ of the compact space $R
\leftrightarrow \alpha'/R$. Here $\alpha'$ is the string slope parameter.
This symmetry is called T-duality. However this duality is not realized manifestly in the theory.

String theory with manifest T-duality has been developed first in \cite{Siegel:1993xq}.
Recently, an effective theory that realizes manifest T-duality has been
proposed \cite{Hull:2009mi}. In this theory, T-duality is manifestly realized by introducing the Fourier
dual of the winding momentum -- the winding coordinate $\tilde{x}_{\mu}$ --
in addition to the ordinary space-time coordinate $x^{\mu}$.
Therefore the space-time is effectively doubled and it is called the doubled space-time.
It was pointed out that the doubled space-time is naturally incorporated
in a para-Hermitian (K\"ahler) manifold \cite{Vaisman:2012ke, Vaisman:2012px}.
An effective supergravity theory defined in the doubled space-time with manifest T-duality
is called doubled field theory (DFT).
Since the dimension of the space-time is doubled, it is necessary to impose a constraint to any quantities in DFT to
obtain a physical space-time.
There are two constraints in DFT \cite{Hohm:2010jy}. 
One is the weak constraint given by $ \partial_{\mu} \tilde{\partial}^{\mu} \Psi = 0 $.
This is equivarent to the level matching condition of closed strings \cite{Hull:2009mi}.
The other is the  strong constraint represented by
$\tilde{\partial}^{\mu} \Psi \partial_{\mu} \Phi + \partial_{\mu} \Psi
\tilde{\partial}^{\mu} \Phi = 0$.
Here $\Psi, \Phi$ are all the quantities (tensor fields
and gauge parameters) defined in the doubled space-time.
The strong constraint is nessesary to close the gauge algebra in DFT.

It is known that DFT exhibits a gauge symmetry that inherits the
diffeomorphism of the tensor fields and the Abelian gauge
symmetry of the $B$-field.
The gauge symmetry of DFT is governed by a bracket, called the {\sf C}-bracket,
which is a generalization of the Courant bracket appearing in the
context of generalized geometry \cite{Courant, LiWeXu}.
The {\sf C}-bracket defines various algebraic structures on the doubled space-time.
Among other things, the Vaisman algebroid (the metric algebroid
in \cite{Vaisman:2012ke, Vaisman:2012px, Svoboda:2018rci,
Carow-Watamura:2020xij} or the pre-DFT algebroid in \cite{Chatzistavrakidis:2018ztm})
is the most natural algebroid appearing in the para-Hermitian manifold.
Under the strong constraint, the {\sf C}-bracket reduces to the Courant bracket and
the Vaisman algebroid reduces to the Courant algebroid \cite{Hull:2009zb}.

In the formulation of DFT, relevant ingredients (space-time, vectors,
tensors, symmetries) are all doubled.
From the mathematical viewpoint,
these doubled structures are shown up as the Drinfel'd double of
algebroids.
Indeed, it is known that the Courant algebroid appearing in the DFT is
obtained by the Drinfel'd double of a Lie bialgebroid \cite{LiWeXu}.
We have shown that this property is inherited to the Vaisman algebroid \cite{Mori:2019slw}.
Given a pair of Lie algebroids with the {\sf C}-bracket and appropriate algebroid structures,
the Vaisman algebroid is obtained by an analogue of the Drinfel'd double.
On the other hand, the Drinfel'd double is a key ingredient of the
Poisson-Lie T-duality (plurality) \cite{Klimcik:1995jn, Klimcik:1995ux,
VonUnge:2002xjf}, and U-duality \cite{Sakatani:2019zrs, Malek:2019xrf,
Blair:2020ndg, Malek:2020hpo}.
The Poisson-Lie T-duality is a generalization of T-duality in string theory defined on a group manifold.
Even though its physical meaning is still mysterious,
it provides various meaningful results including solution generating
technique \cite{Hassler:2017yza, Demulder:2018lmj , Sakatani:2019jgu, Hlavaty:2019pze, Musaev:2020bwm},
integrable deformations of string theory and applications to the AdS/CFT correspondence \cite{Delduc:2013qra}.
The Poisson-Lie T-duality is based on a choice of the Manin triple decompositions behind the Drinfel'd double for
a group manifold $G$.
A Manin triple is defined for a Lie bialgebra associated with a Poisson-Lie group $G$.
Therefore the doubled nature of algebras (algebroids) is an essential feature of T-duality and its generalizations.

In this paper, we study doubled aspects of algebroids behind DFT.
As we will discuss below, there are several possible algebroid structures
which encompasses the doubled nature and the {\sf C}-bracket in DFT.
We stress that our analysis can be applied to DFT on curved or group manifolds $\mathcal{M}$.
When the underling manifold is a Lie group, one can define globally
defined left-invariant vector fields on $\mathcal{M}$.
The Lie algebra of this left-invariant vector fields is isomorphic to
the Lie algebra $\mathfrak{g}$ of $G$.
Indeed, if $G$ is a Drinfel'd double, then, $\mathfrak{g}$ becomes a Lie bialgebra.

The organization of this paper is as follows.
In the next section, we present a brief overview of the Courant algebroid and its doubled structure.
We also make a brief overview our previous work \cite{Mori:2019slw} where the doubled structure of
the Vaisman algebroid was discussed.
We then classify a family of
algebroids equipped with the {\sf C}-bracket and those respect the doubled structure.
They originate from the Vaisman algebroid and settle down to the
Courant algebroid under the appropriate conditions.
In Section \ref{sect:DFT-algebroids}, we then study the doubled aspects of these algebroids.
We study consistency conditions for these algebroids step by step.
A careful analysis reveals that weaker versions of the Courant
algebroids, known as the pre- and the ante-Courant algebroids discussed
in \cite{Chatzistavrakidis:2018ztm}
are build out of Lie algebroid pairs.
We show that the doubled nature of the pre-Courant algebroid allows non-trivial Poisson structures.
In Section \ref{sect:DFT}, we study geometric realizations of these algebroids in DFT.
We consider the para-Hermitian manifold as a natural arena of the doubled space-time.
We then show that the various algebroid structures arise in the tangent
bundle of the para-Hermitian manifold.
In the implementation of these algebroids in the para-Hermitian
manifold, we find that relaxed versions of the strong constraint is
necessary for the doubled structures.
We also study the consistency conditions of fluxes and doubled structures in the para-Hermitian manifold.
In Section \ref{sect:group}, we briefly discuss the doubled structures in a group manifold.
Section \ref{sect:conclusion} is the conclusion and discussions.

\section{Courant and Vaisman algebroids and their doubled structures} \label{sect:VC-algebroids}
In this section, we introduce the Courant and the Vaisman algebroids
that appear in DFT.
We then classify several kinds of algebroids equipped with the {\sf C}-bracket.

We first introduce Lie algebroids.
Let $E$ be a vector bundle over a manifold $M$.
A skew-symmetric bracket $[\cdot,\cdot]_E: \Gamma(E) \times \Gamma(E)
\rightarrow \Gamma(E)$ that satisfies the Jacobi identity is defined.
We also introduce a bundle map called the anchor $\rho_E: E \rightarrow TM$.
Then, a Lie algebroid $E$ is defined by $(E,[\cdot,\cdot]_E,\rho_E)$.
By the dual vector bundle $E^*$, we can also define the dual Lie
algebroid $(E^*, [\cdot,\cdot]_{E^*}, \rho_{E^*})$.

We are interested in the algebraic structure defined by the following {\sf C}-bracket:
\begin{align}
  [e_1,e_2]_{\sf C}
  &= [X_1,X_2]_E + \mathcal{L}_{\xi_1}X_2 - \mathcal{L}_{\xi_2}X_1 -
 {\rm d}_* \bbra{e_1,e_2}_- \notag\\
  \label{def:Vaisman_bracket}
  &\quad  + [\xi_1,\xi_2]_{E^*} + \mathcal{L}_{X_1}\xi_2 -
 \mathcal{L}_{X_2}\xi_1 + {\rm d} \bbra{e_1,e_2}_-.
\end{align}
where $X_i\in\Gamma(E), \xi_i\in\Gamma(E^*), e_i = X_i + \xi_1
\in\Gamma(E\oplus E^*)$, $\dop, \dop_*$ and
$\mathcal{L}_{X}, \mathcal{L}_{\xi}$ are the exterior derivatives and
the Lie derivatives on $\Gamma(\wedge^p E)$, $\Gamma(\wedge^p E^*)$,
respectively.
A non-degenerate bilinear form $\bbra{e_1,e_2}_-$, whose explicit form will
be given later, has been introduced.
The {\sf C}-bracket governs the $O(D,D)$ T-duality covariant gauge symmetry in DFT
\cite{Hull:2009mi}.

The Vaisman algebroid is one of the notable structure defined by the
 {\sf C}-bracket.
This algebroid has initially been introduced as a metric algebroid in
 \cite{Vaisman:2012ke,Vaisman:2012px}.
The pre-DFT algebroid discussed in \cite{Chatzistavrakidis:2018ztm} also refers to the same algebroid.
The Vaisman algebroid is defined by the following structures.
Let $\mathcal{V}$ be a vector bundle over $M$.
We introduce a skew-symmetric bracket $[\cdot,\cdot]_{\rm
V}:\Gamma(\mathcal{V}) \times \Gamma(\mathcal{V}) \to
\Gamma(\mathcal{V}) $ called the Vaisman bracket.
The anchor $\rho_{\rm V}$ is defined by a bundle map $\mathcal{V} \to
TM$.
In addition, we introduce a symmetric bilinear form $\bbra{\cdot,\cdot}:
\mathcal{V} \times \mathcal{V} \to \mathbb{R}$.
Then, $(\mathcal{V}, [\cdot,\cdot]_{\rm V}, \rho_{\rm V},
\bbra{\cdot,\cdot})$ defines a Vaisman algebroid if the following two
axioms are satisfied.
\begin{axiomV}
\label{axiom:V1}
For any $e_1, e_2 \in \Gamma (\mathcal{V})$, $f \in C^{\infty} (M)$,
\begin{align}
[e_1, f e_2]_{\rm V} = f [e_1, e_2]_{\rm V} + (\rho_{\rm V} (e_1) \cdot f) e_2 - \bbra{e_1, e_2}
 \mathcal{D}f.
\label{eq:V1}
\end{align}
where $\mathcal{D}$ is a map from $C^\infty(M) \to \Gamma(\mathcal{V})$.
\end{axiomV}
\begin{axiomV}
\label{axiom:V2}
For any $e_1, e_2, e_3 \in \Gamma (\mathcal{V})$, we have the
 compatibility condition between the bilinear form $\bbra{\cdot,\cdot}$ and the anchor $\rho_{\rm V}$:
\begin{align}
\rho_{\rm V} (e_1) \cdot \bbra{e_2,e_3} = \bbra{[e_1, e_2]_{\rm V} + \mathcal{D} \bbra{e_1,e_2}, e_3}
+\bbra{e_2, [e_1, e_3]_{\rm V} + \mathcal{D} \bbra{e_1,e_3}}.
\label{eq:V2}
\end{align}
\end{axiomV}

In the following, we show that
a Vaisman algebroid equipped with the {\sf C}-bracket is constructed
by the Lie algebroid pair $E$ and $E^*$ \cite{Mori:2019slw}.
Sections on the Lie algebroids $E,E^*$ are denoted by $X_i \in \Gamma(E)$
and $\xi_i \in \Gamma(E^*)$, respectively.
Next, we define a new vector bundle $\mathcal{V}= E\oplus E^*$ over $M$.
Then an element in $\Gamma(\mathcal{V})$ is denoted by $e_i= X_i + \xi_i$.
We define the non-degenerate bilinear forms $\pmbra{\cdot,\cdot}$ on
$\Gamma(\mathcal{V})$ by the inner product $\langle \cdot, \cdot \rangle
$ between $E$ and $E^*$:
\begin{equation}
  \pmbra{e_1,e_2} = \frac{1}{2}(\langle \xi_1,X_2 \rangle \pm \langle\xi_2,X_1 \rangle).
\end{equation}
We define the anchor $\rho_{\rm V}$ as
\begin{equation}
  \rho_{\rm V}(e_i) = \rho_E(X_i) + \rho_{E^*}(\xi_i),
\end{equation}
where $\rho_E, \rho_{E^*}$ are the anchors of the Lie algebroids $E,E^*$.
We finally introduce the map $\mathcal{D}$ as
\begin{equation}
  \pbra{\mathcal{D}f,e_i} = \frac{1}{2}\rho_{\rm V}(e_i)f, \quad f \in
   C^{\infty} (M).
  \label{eq:realizeD}
\end{equation}
Together with the definition of $\rho_{\rm V}$ above, we can denote
$\mathcal{D} = {\rm d} + {\rm d}_*$, where $\dop, \dop_*$ are exterior
derivatives on $E^*$ and $E$.
With these definitions at hand, we now provide a brief summary of the proof for
the Axioms V1 and V2 (see \cite{Mori:2019slw} for details).
First, we confirm the condition \eqref{eq:V1} for Axiom V1.
Using the explicit form of the {\sf C}-bracket
\eqref{def:Vaisman_bracket} and the general properties of the Lie
derivatives $\mathcal{L}_{X_1}, \mathcal{L}_{\xi_i}$ \cite{Mackenzie},
the left-hand side of the equation \eqref{eq:V1} is expanded and
evaluated as
\begin{align}
  [X_1, fX_2]_{\sf C}
  & = f[X_1,X_2]_{\sf C} + (\rho_E(X_1) \cdot f ) X_2, \notag\\
  [X_1, f\xi_2]_{\sf C}
  & = f[X_1,\xi_2]_{\sf C} + (\rho_E (X_1) \cdot f) \xi_2 - \frac{1}{2} \mathcal{D} f\langle \xi_2,X_1 \rangle, \notag\\
  [\xi_1, fX_2]_{\sf C}
  & = f[\xi_1,X_2]_{\sf C} + (\rho_{E^*} (\xi_1) \cdot f) X_2 - \frac{1}{2}\mathcal{D} f \langle \xi_1,X_2 \rangle,\notag\\
  [\xi_1,f\xi_2]_{\sf C}
  \label{eq:part_of_V1}
  &= f [\xi_1,\xi_2]_{\sf C} + (\rho_{E^*} (\xi_1) \cdot f) \xi_2.
\end{align}
It is obvious that the condition \eqref{eq:V1} is satisfied by summing up all the four equations in \eqref{eq:part_of_V1}.
Therefore, Axiom V1 holds by the structure $(\mathcal{V},[\cdot,\cdot]_{\sf C},\rho_{\rm V},\pbra{\cdot,\cdot})$.

Next, we confirm Axiom V2 defined by the equation \eqref{eq:V2}.
Since the Vaisman bracket is given by the {\sf C}-bracket, we can show the following relation by the direct calculations:
\begin{align}
\pbra{[e_1,e_2]_{\rm V},e_3} =& \  T(e_1,e_2,e_3) + \frac{1}{2} \rho_{\rm V} (e_1) \cdot
\pbra{e_2,e_3}
\notag \\
& \ - \frac{1}{2} \rho_{\rm V} (e_2) \cdot \pbra{e_1,e_3}
\label{eq:V5sum}
\pbra{e_3,e_2} - \frac{1}{2} \rho_{\rm V} (e_3) \cdot \pbra{e_1,e_2},
\end{align}
where $T(e_1,e_2,e_3)$ is defined by
\begin{equation}
  T(e_1,e_2,e_3) = \frac{1}{3}
\Bigl(
\pbra{[e_1,e_2]_{\rm V},e_3} + \text{c.p.}
\Bigr).
\end{equation}
Here, c.p. stands for terms with the cyclic permutations of 1,2,3.
By the definition of the anchor $\rho_{\rm V}$ in \eqref{eq:realizeD}, we have
\begin{align}
  \frac{1}{2}\rho_{\rm V} (e_1) \cdot \pbra{e_3,e_2}
&= \pbra{\mathcal{D}\pbra{e_3,e_2} , e_1}.
\label{eq:rho_deriv}
\end{align}
With these results, one finds that the condition \eqref{eq:V2} holds.
Therefore, the quadruple $(\mathcal{V}, [\cdot,\cdot]_{\sf C}$, $\rho_{\rm
V}, \pbra{\cdot,\cdot})$ satisfies Axioms V1, V2 and define a Vaisman algebroid.
Since we introduce $E\oplus E^*$ as the vector bundle $\mathcal{V}$, we
call this procedure the ``double'' of $E$ and $E^*$.

On the other hand, there is another algebraic structure described by the
{\sf C}-bracket. This is known as the Courant algebroid.
This is defined as follows.
Let $\mathcal{C}$ be a vector bundle over $M$.
We define a skew-symmetric bracket $[\cdot,\cdot]_{\rm c}:
\Gamma(\mathcal{C}) \times \Gamma(\mathcal{C}) \to \Gamma(\mathcal{C})$
called the Courant bracket and a bilinear form $\bbra{\cdot,\cdot}:
\Gamma(\mathcal{C}) \times \Gamma(\mathcal{C}) \to \mathbb{R}$.
The set of these structures $(\mathcal{C},[\cdot,\cdot]_{\rm
c},\rho_{\rm c},\bbra{\cdot,\cdot})$
becomes a Courant algebroid if the following axioms are satisfied.

\begin{axiomC}
\label{axiom:C1}
For any $e_1, e_2, e_3 \in \Gamma (\mathcal{C})$, the Jacobiator of
      $[\cdot,\cdot]_{\rm c}$ is given by
\begin{align}
[[e_1, e_2]_{\rm c}, e_3]_{\rm c} + \text{c.p.} = \mathcal{D} T (e_1, e_2, e_3),
\label{eq:C1}
\end{align}
 where $\mathcal{D}$ is the map $C^\infty(M) \to \Gamma(\mathcal{C})$,
 $T (e_1, e_2, e_3) = \frac{1}{3} \bbra{ [e_1, e_2]_{\rm c}, e_3 } + \text{c.p.}$
 and $\text{c.p.}$ is the
 cyclic permutations.
\end{axiomC}
\begin{axiomC}
\label{axiom:C2}
For any $e_1, e_2 \in \Gamma (\mathcal{C})$,
\begin{align}
\rho_{\rm c} ([e_1, e_2]_{\rm c}) =  [\rho_{\rm c} (e_1), \rho_{\rm c} (e_2)].
\label{eq:C2}
\end{align}
Here, $[\cdot,\cdot]$ is the Lie bracket on $TM$.
\end{axiomC}
\begin{axiomC}
\label{axiom:C3}
For any $e_1, e_2 \in \Gamma (\mathcal{C})$, $f \in C^{\infty} (M)$,
\begin{align}
[e_1, f e_2]_{\rm c} = f [e_1, e_2]_{\rm c} + (\rho_{\rm c} (e_1) \cdot f) e_2 - \bbra{e_1, e_2}
 \mathcal{D}f.
\label{eq:C3}
\end{align}
\end{axiomC}
\begin{axiomC}
\label{axiom:C4}
$\rho_{\rm c} \cdot \mathcal{D} = 0$, namely, for any $ f,g \in
      C^{\infty} (M)$, we have
\begin{align}
\bbra{\mathcal{D}f, \mathcal{D}g} = 0.
\label{eq:C4}
\end{align}
\end{axiomC}
\begin{axiomC}
\label{axiom:C5}
We have the compatibility condition
 between the bilinear form $\bbra{\cdot,\cdot}$ and the anchor $\rho_{\rm c}$ as follows.
 For any $e_1, e_2, e_3 \in \Gamma (\mathcal{C})$,
\begin{align}
\rho_{\rm c} (e_1) \cdot \bbra{e_2,e_3} = \bbra{[e_1, e_2]_{\rm c} + \mathcal{D} \bbra{e_1,e_2}, e_3}
+\bbra{e_2, [e_1, e_3]_{\rm c} + \mathcal{D} \bbra{e_1,e_3}}.
\label{eq:C5}
\end{align}
\end{axiomC}

One finds that Axioms C3 and C5 correspond to Axioms V1 and V2 respectively.
In this sense, any Courant algebroids are Vaisman algebroids
but the converse is not true.
Indeed, the Vaisman algebroid $(\mathcal{V},[\cdot,\cdot]_{\sf C},\rho_{\rm
 V},\pbra{\cdot,\cdot})$ made from the double above does not have a
 Courant algebroid structure in general.
The Jacobiator of the {\sf C}-bracket is calculated as
\begin{align}
  [[e_1,e_2]_{\sf C},e_3]_{\sf C} + \text{c.p.} =
 \mathcal{D} T(e_1,e_2,e_3) - ( J_1 + J_2 + \text{c.p.}).
\label{eq:Jac_V}
\end{align}
where $J_1, J_2$ have been defined by
\begin{align}
  J_1
  &= \iop_{X_3}
\Bigl(
{\dop}[\xi_1,\xi_2]_{E^*} - \mathcal{L}_{\xi_1}{\dop}\xi_2 +
 \mathcal{L}_{\xi_2}{\dop}\xi_1
\Bigr)
+
\iop_{\xi_3}
\Bigl(
{\dop}_*[X_1,X_2]_E - \mathcal{L}_{X_1}{\dop}_*X_2 +
 \mathcal{L}_{X_2}{\dop}_*X_1
\Bigr),
\notag\\
  J_2
  &=
\Bigl(
\mathcal{L}_{{\dop}_* \mbra{e_1,e_2}}\xi_3 + [{\dop}\mbra{e_1,e_2},\xi_3]_{E^*}
\Bigr)
 -
\Bigl(
\mathcal{L}_{{\dop} \mbra{e_1,e_2}}X_3 + [{\dop}_*\mbra{e_1,e_2},X_3]_E
\Bigr).
\label{eq:J1J2}
\end{align}
The terms $J_1$ and $J_2$ do not vanish in general.
Therefore, Axiom C1 is broken for the Vaisman algebroid
$(\mathcal{V},[\cdot,\cdot]_{\sf C},\rho_{\rm V},\pbra{\cdot,\cdot})$.
However, one can show that $J_1$ and $J_2$ vanish by imposing the
following derivation condition for $\dop_*$ for
all $X, Y \in \Gamma (E)$ \cite{LiWeXu}:
\begin{align}
\dop_{*} [X,Y]_{\rm S} = [\dop_* X, Y]_{\rm S} + [X,\dop_* Y]_{\rm S}.
\label{eq:derivation}
\end{align}
Here, $[\cdot,\cdot]_{\rm S}$
is the Schouten-Nijenhuis bracket
on $\Gamma(\wedge^p E)$ \cite{Mackenzie}.
Similarly, one can confirm that Axioms C2 and C4 follow from the
condition \eqref{eq:derivation} \cite{Mori:2019slw}.
Therefore, the Vaisman algebroid $(\mathcal{V},[\cdot,\cdot]_{\sf
C},\rho_{\rm V},\pbra{\cdot,\cdot})$ becomes a Courant algebroid when
the derivation condition \eqref{eq:derivation} is imposed.
It has been discussed that the Courant algebroid
$(\mathcal{V},[\cdot,\cdot]_{\sf C},\rho_{\rm V},\pbra{\cdot,\cdot})$
is constructed by the Drinfel'd double of a Lie bialgebroid \cite{LiWeXu}.
The Lie bialgebroid is defined by a pair of dual Lie algebroids
$(E,E^*)$ together with the derivation condition \eqref{eq:derivation}.
When a pair $(E,E^*)$ is a Lie bialgebroid, $(E^*,E)$ also becomes a Lie bialgebroid.
Then, as with \eqref{eq:derivation}, the following relation also holds:
\begin{align}
\dop [\xi,\eta]_{\rm S^*} = [\dop \xi, \eta]_{\rm S^*} + [\xi,\dop \eta]_{\rm S^*}.
\label{eq:derivation_dual}
\end{align}

Now, we are interested in what kind of algebroids equipped with the {\sf
C}-bracket are allowed other than the Vaisman and the Courant algebroids.
As we have shown above, the derivation condition \eqref{eq:derivation}
is necessary for the Courant algebroid and it
implies the strong constraint in DFT
in a specific geometric realization \cite{Mori:2019slw}.
However it seems to be too strong.
Indeed, teh derivation condition \eqref{eq:derivation} is a sufficient condition for the strong constraint.
We would like to study whether the relaxation of the strong constraint
is possible or not by examining the intermediate structures between the
Vaisman and the Courant algebroids.

\begin{figure}[t]
  \centering
  \includegraphics[width=12cm]{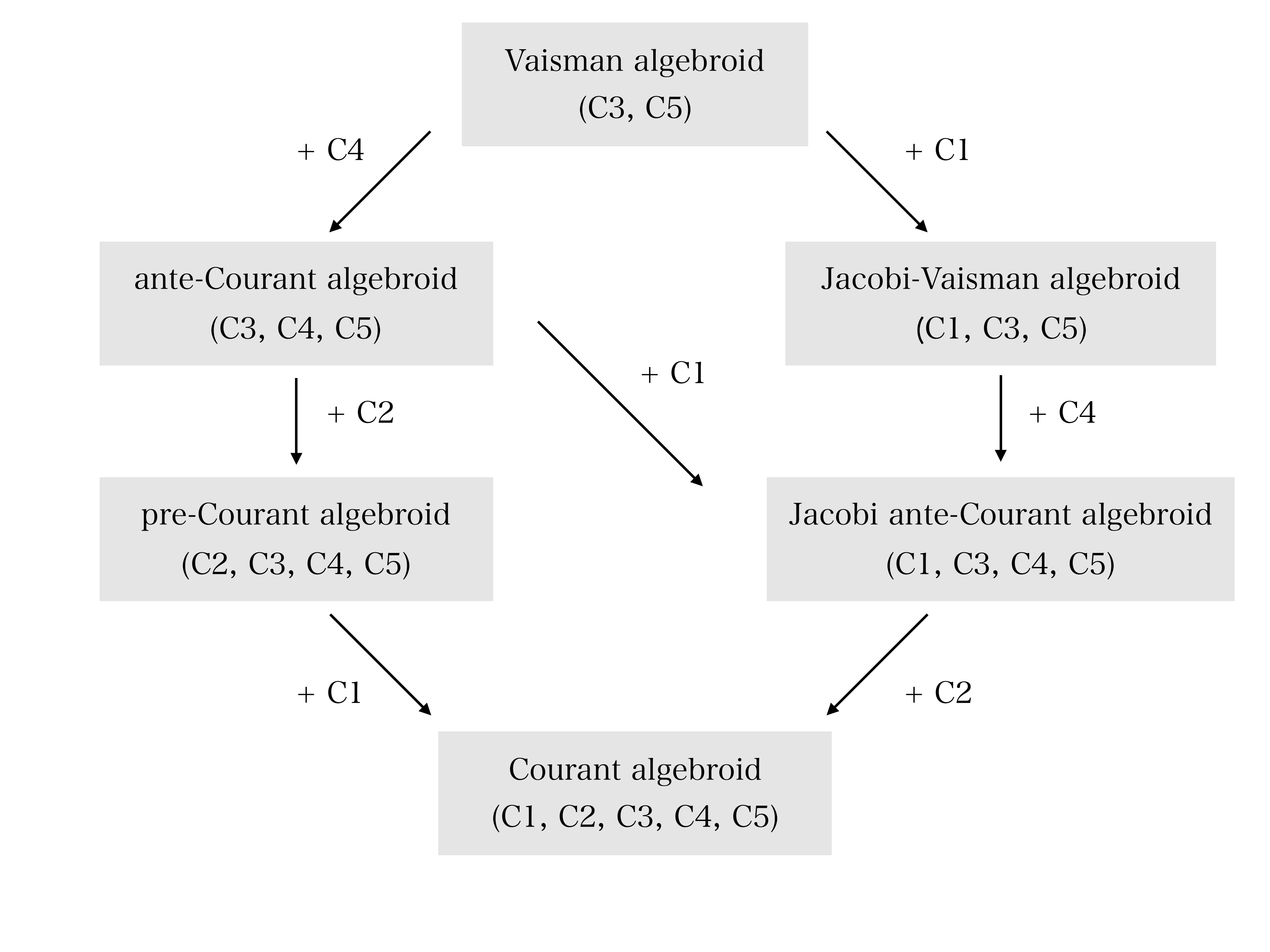}
  \caption{
  A list of algebroids allowed by the combinations of Axioms C1- C5.
  They are classified into the two sequences according to the logical structures of axioms.
}
  \label{fig.1}
\end{figure}

First, we focus on each Axiom and write down all the possible algebroid
structures that with the {\sf C}-bracket.
In general, Axioms C1-C5 of a Courant algebroid are not independent with
each other \cite{Uchino:2002}.
Indeed, Axiom C5 implies C3, and C2 implies C4.
Only Axiom C1 is independent of the other Axioms.
Therefore, the possible combinations of Axioms in the general case are
found to be
\begin{align}
  &(C1),\ \ (C3),\ \ (C4),\ \ (C2, C4),\ \ (C3,C5), \notag\\
  &(C1, C3),\ \ (C1, C4),\ \ (C3,C4),\ \ \notag\\
  &(C1,C2, C4),\ \ (C1, C3, C4),\ \ (C1, C3, C5),\ \ (C2, C3, C4),\ \ (C3, C4, C5), \notag\\
  &(C1,C2, C3,C4),\ \ (C1, C3, C4, C5),\ \ (C2, C3, C4, C5),\ \ \notag\\
  &(C1, C2, C3, C4, C5).
  \label{classifi_alg}
\end{align}
However, as we have seen above, an algebroid made by the double together
with the {\sf C}-bracket \eqref{def:Vaisman_bracket} necessarily satisfies
Axioms C3 and C5.
Thus the possible combinations of \eqref{classifi_alg} reduces to the
following ones:
\begin{align}
  &(C3,C5), \ \ (C1, C3,C5),\ \ (C3,C4,C5),\ \
  \notag\\
  &(C1, C3, C4, C5),\ \ (C2, C3, C4, C5),\ \ (C1, C2, C3, C4, C5).
  \label{classifi_alg_2}
\end{align}
In \eqref{classifi_alg_2}, $(C1, C2, C3, C4, C5)$ and $(C3,C5)$
correspond to the Courant and the Vaisman algebroids, respectively.
An algebroid defined by $(C2, C3, C4, C5)$ is known to be
the pre-Courant algebroid \cite{Vaisman:2004}.
An algebroid by $(C3, C4, C5)$ has been introduced in
\cite{Chatzistavrakidis:2018ztm} and is called the ante-Courant
algebroid.
On the other hand, the other possibilities $(C1, C3, C5)$, $(C1, C3, C4,
C5)$ have not been discussed in the literature.
Since Axiom C1 means the (modified) Jacobi identity,
we call $(C1, C3,C5)$ the Jacobi Vaisman algebroid while $(C1, C3, C4,
C5)$ the Jacobi ante-Courant algebroid.
Note that, this ``Jacobi'' is not related to the Jacobi structure
proposed by Lichnerowicz which is a generalization of the Poisson structure.
All of these algebroids are summarized in Figure \ref{fig.1}.
We collectively call these the DFT algebroids.
They are organized into two parts.
One is those with the Jacobi identity corresponding to the
right flow in Figure \ref{fig.1}.
The other is the ones without the Jacobi identity in the left flow.

In the next section, we examine the compatibility conditions between the
{\sf C}-bracket and the doubled structures in the algebroids classified above.

\section{Doubled structures of algebroids with {\sf C}-bracket} \label{sect:DFT-algebroids}

In this section, we study doubled structures of algebroids defined by the
{\sf C}-bracket.
We focus on the Jacobi Vaisman, the Jacobi
ante-Courant, the ante-Courant and the pre-Courant algebroids.

\subsection{Doubled structures of Jacobi Vaisman and Jacobi ante-Courant algebroids}

We first consider the Jacobi Vaisman algebroid in the series of the
 right flow in Figure \ref{fig.1}.
The Jacobi Vaisman algebroid is obtained by imposing Axiom C1 to a
 Vaisman algebroid. We start from the Vaisman algebroid
 $(\mathcal{V},[\cdot,\cdot]_{\sf C},\rho_{\rm V},\pbra{\cdot,\cdot})$
 made by a pair of Lie algebroids $(E,E^*)$ and impose Axiom C1 to that.
As we have seen before, Axiom C1 needs conditions $J_1 = J_2 = 0$.
The condition $J_1 = 0$ is nothing but the derivation condition \eqref{eq:derivation} itself.
On the other hand, if we assume the derivation condition \eqref{eq:derivation}, we find
\begin{align}
0 =& \
- f
\Bigl(
\dop_* [X,Y]_E + \mathcal{L}_{Y} \dop_* X - \mathcal{L}_{X} \dop_* Y
\Bigr) +
\Bigl(
\dop_* [X,f Y]_E - \mathcal{L}_X \dop_* (f Y) + \mathcal{L}_{fY} \dop_* X
\Bigr)
\notag \\
=& \
\left(
\mathcal{L}_{\dop f} X + [\dop_* f,X]_E
\right) \wedge Y
\label{eq:C4_V}
\end{align}
for any $X,Y \in \Gamma(E)$.
A similar result holds for \eqref{eq:derivation_dual}.
Therefore, we obtain the following relations
\begin{align}
 \mathcal{L}_{\dop f} X = - [\dop_* f, X]_L, \qquad
 \mathcal{L}_{\dop_* f} \xi = - [\dop f, \xi]_{\tilde{L}}.
  \label{eq:prop3.4_v}
\end{align}
If we consider $ X = \dop_* \pbra{e_1,e_2}, \xi = \dop
\pbra{e_1,e_2}$ in equation \eqref{eq:prop3.4_v}, we have $J_2 + {\rm c.p.}=0$.
Therefore, it is sufficient to impose the derivation condition to make
$J_1 + {\rm c.p.} =0$ and $J_2 + {\rm c.p.}=0$.
Therefore, the derivation condition is the only necessary condition for
the Jacobi Vaisman algebroid.
However, as we will see in below, this condition induces Axioms C2 and C4.
Let us first examine Axiom C4.
Given the Vaisman algebroid $(\mathcal{V},[\cdot,\cdot]_{\sf C},\rho_{\rm
V},\pbra{\cdot,\cdot})$, we find the equation \eqref{eq:C4} is evaluated
as
\begin{align}
    \pbra{\mathcal{D}f,\mathcal{D}g}
    &= \frac{1}{2}( \langle \dop f,\dop_* g \rangle  + \langle \dop_* f, \dop g \rangle) \notag\\
    &= \frac{1}{2}(\rho_{E^*}\rho_{E}^{*} + \rho_{E}\rho_{E^{*}}^{*}){\rm d}_0f \cdot g,
\quad f, g \in C^{\infty} (M).
    \label{eq:C4deform}
\end{align}
Here we have used the relation $\dop_* = \rho_{E^{*}}^* \dop_0, \dop = \rho_{E^{*}}
\dop_0$ and $\dop_0$ is the ordinary exterior derivative on $T^*M$.
Note that the superscript $\bullet^*$ on the anchor means the adjoint operator,
which is defined by the transposition of the original operator
through the inner product $\langle \cdot, \cdot \rangle$.
Namely, it is defined by $\langle \xi, \rho_{E} (X) \rangle = \langle
\rho_{E}^{*} (\xi), X \rangle$ for any $X \in \Gamma(E), \xi \in \Gamma(T^{*} M)$.
The following is the summary of the anchor structures:
\begin{align}
  &\rho_{E}:E\rightarrow TM,\ \ \ \ \ \rho_E^*:T^*M\rightarrow E^*, \notag\\
  &\rho_{E^{*}}:E^{*} \rightarrow TM,\ \ \ \rho^{*}_{E^{*}}:T^{*} M \rightarrow E.
\end{align}
The rightmost side of the equation \eqref{eq:C4deform} seems to be
generally non-zero.
However,
as we have clarified, the derivation condition \eqref{eq:derivation}
induces the condition \eqref{eq:prop3.4_v}.
If we consider $ X = \dop f$ in \eqref{eq:prop3.4_v}, we obtain
\begin{align}
\dop_*
\Bigl(
\rho_{E} \rho_{E^*}^* (\dop_0 f) \cdot f
\Bigr)  = 0, \qquad {}^{\forall} f \in C^{\infty} (M).
\label{eq:PP34-1}
\end{align}
This means
\begin{align}
0 = \rho_{E} \rho_{E^*}^*({\rm d}_0f)\cdot f =
\langle \rho_{E} \rho_{E^*}^*({\rm d}_0f),{\rm d}_0f \rangle.
\label{C4_2}
\end{align}
The result \eqref{C4_2} is equivalent to the condition that the right-hand side of
\eqref{eq:C4deform} vanishes. This also means the anti-symmetric
property of the anchor map:
\begin{align}
  \label{eq:anchor_sym}
  \rho_{E} \rho_{E^{*}}^{*} = - \rho_{E^*} \rho_E^*.
\end{align}
Therefore Axiom C4 is automatically satisfied by imposing the derivation
condition \eqref{eq:derivation}.

Next, we clarify Axiom C2.
Given the Vaisman algebroid $(\mathcal{V},[\cdot,\cdot]_{\sf
C},\rho_{\rm V},\pbra{\cdot,\cdot})$, we calculate
the difference between the two sides of the equation \eqref{eq:C2}.
The result is
\begin{align}
&
\rho_{\rm V} ([e_1,e_2]_{\rm V}) \cdot f -
[\rho_{\rm V} (e_1),\rho_{\rm V} (e_2)] f
\notag \\
& =
 - \langle \xi_1 , \bigl(\mathcal{L}_{{\dop}f}X_2 - [X_2,{\dop}_*f]_E \bigr) \rangle
+ \langle \xi_2 , \bigl(\mathcal{L}_{{\dop} f}X_1 - [X_1,{\dop}_*f]_E
 \bigr) \rangle\notag\\
& \quad \,
+ \frac{1}{2}
\Bigl(
\rho_E \rho_{E^*}^*
+
\rho_{E^*} \rho_E^*
\Bigr)
{\dop}_0(\langle \xi_1,X_2 \rangle
- \langle \xi_2,X_1 \rangle ) \cdot f.
 \label{C2_7}
\end{align}
It is obvious that the right-hand side of \eqref{C2_7} vanishes by
the conditions \eqref{eq:prop3.4_v} and \eqref{eq:anchor_sym} that are induced by \eqref{eq:derivation}.
Then, Axiom C2 is automatically satisfied due to the derivation condition.

In summary, as long as we employ the {\sf
C}-bracket and the doubled structure,
it is impossible to construct the Jacobi Vaisman algebroid
that satisfies only Axioms C1, C3, C5.
The same is true even for the Jacobi ante-Courant algebroid.
However, we stress that the result in this section does not mean the no-go theorem of these algebroids.
Our discussion critically depends on the doubled structure of the {\sf C}-bracket.
If we do not persist on the doubled structure or the {\sf C}-bracket,
there are still room for these algebroids by defining a suitable bracket, instead of the {\sf C}-bracket,  that satisfies appropriate axioms.

\subsection{Doubled structures of ante- and pre-Courant algebroids}

Next, we consider the series of the left flow in Figure \ref{fig.1}.
We will clarify the compatibility conditions between the doubled
structures and the {\sf C}-bracket for these algebroids.
Compared with the Jacobi Vaisman and the Jacobi ante-Courant algebroids,
Axiom C1 is not required for ante- and pre-Courant algebroids.

First, we consider the ante-Courant algebroid.
This is obtained by imposing Axiom C4 to a Vaisman algebroid.
Again, we consider the Vaisman algebroid
 $(\mathcal{V},[\cdot,\cdot]_{\sf C},\rho_{\rm V},\pbra{\cdot,\cdot})$
 made by a pair of Lie algebroids $(E,E^*)$.
The only condition that we need for ante-Courant algebroids
is the anti-symmetric nature of the anchor \eqref{eq:anchor_sym}.
In the previous section, we showed that \eqref{eq:anchor_sym} is induced
by the derivation condition but the converse is not true.
Therefore, even though the condition \eqref{eq:anchor_sym} is imposed,
this does not imply the Jacobi identity and Axiom C1.
The same is true for Axiom C2.

For the pre-Courant algebroid, we need to impose Axiom C2 in addition to
 C4 to the Vaisman algebroid.
From the discussion in the previous section, the condition
 \eqref{eq:prop3.4_v} for Axiom C2 implies the anti-symmetric nature of
 the anchor \eqref{eq:anchor_sym} required by Axiom C4.
Then we conclude that only the condition for the pre-Courant algebroid
 is the equation \eqref{eq:prop3.4_v}.

A comment on the condition \eqref{eq:prop3.4_v} is in order.
When we take $X = \dop_* g$ in the first equation in
\eqref{eq:prop3.4_v}, we have
\begin{align}
[\dop_{*} g, \dop_{*} f] = \mathcal{L}_{\dop f} \dop_{*} g = \dop_{*}
 (\iota_{\dop f} \dop_{*} g), \quad f,g \in C^{\infty} (M).
\end{align}
Since $\iota_{\dop f} \dop_{*} g = \langle \dop f, \dop_{*} g \rangle =
\langle \dop_0 f, \rho_{E} \rho^{*}_{E^{*}} (\dop_0 g) \rangle$, we find
\begin{align}
[\dop_{*} g, \dop_{*} f] =
\dop_{*}
\Big(
 \rho_{E} \rho^{*}_{E^{*}} (\dop_0 g) [f]
\Big).
\label{eq:poisson}
\end{align}
Here we have used the notation $X[f] = \langle X, \dop_0 f \rangle$ for vectors
$X \in \Gamma (TM)$. Now we define a structure $\{ g,f \}$ by
\begin{align}
\{g, f \} = \pi (\dop_{*} g) [f],
\end{align}
where $\pi =  \rho_{E} \rho^{*}_{E^{*}}$.
It is easy to show that this structure is skew-symmetric and possesses
the bilinear nature. Since $\pi (\dop_{*} g)$ belongs to
$\Gamma (TM)$, the operator $\{g, \cdot\}$ acts on functions as a
derivation.
Furthermore, by acting $\rho_E$ on the both sides of the relation \eqref{eq:poisson},
one can show that
\begin{align}
\pi \dop_0
\left(
\Big\{
\{
g,f
\},
h
\Big\}
\right)
=
\Big[
[
\pi \dop_0 g, \pi \dop_0 f
]
,
\pi \dop_0 h
\Big],
\quad f,g,h \in C^{\infty} (M).
\end{align}
Since the right-hand side is given by the Lie bracket, the structure
$\{\{g,f\}, h \}$ satisfies the Jacobi identity.
These properties are enough to conclude that $\{g,f\}$ provides a
non-trivial Poisson structure in $M$. One finds that $\bar{\pi} =
\rho_{E^{*}} \rho_{E}^{*}$ also defines another Poisson structure.
Although, this result was discussed first in \cite{Mackenzie} in the
context of Lie bialgebroids, we stress that the essential property is
\eqref{eq:prop3.4_v} and the condition for the pre-Courant algebroid is
necessary to define non-trivial Poisson structures in $M$.

In the next section, we write down the geometric expressions for these
conditions in the explicit realization of the doubled geometry in DFT.

\section{DFT realization of algebroids in para-Hermitian manifold} \label{sect:DFT}

In double field theory, we need two types of coordinates to describe the
doubled space-time.
One is $x^\mu$ which is the Fourier dual of the KK modes.
The other is $\tilde{x}_\mu$ which is the Fourier dual of the winding
modes of strings.
We denote these as $X^M = (x^\mu,\tilde{x}_\mu), \ (M=1, \ldots 2D, \mu =
1, \ldots D)$.
The structure of the doubled space-time appears naturally in the
$2D$-dimensional flat para-Hermitian manifold
\cite{Vaisman:2012ke, Vaisman:2012px, Freidel:2017yuv, Freidel:2018tkj, Marotta:2018myj}.
In our previous paper \cite{Mori:2019slw},
we realized the Vaisman and the Courant algebroids with the doubled
structure on a para-Hermitian manifold.
We showed that the derivation condition \eqref{eq:derivation}
corresponds to the strong constraint of DFT in this setup.

Now, we are interested in whether or not the strong constraint can be relaxed.
In the previous section, we derive the conditions for the ante- and the pre-Courant
algebroids equipped with the {\sf C}-bracket and the doubled structure.
Since these conditions are induced by the derivation condition, but the
converse is not true, we can relax the strong constraint in the ante-
and the pre-Courant algebroids.

In this section, we first introduce the general para-Hermitian manifold.
Then, we construct the ante- and the pre-Courant algebroid and write
down the conditions \eqref{eq:prop3.4_v} and
\eqref{eq:anchor_sym} in the geometric form on the flat para-Hermitian manifold.

\subsection{Para-Hermitian manifold}
A para-Hermitian manifold $\mathcal{M}$ is defined as follows.
Let $\mathcal{M}$ be a $2D$-dimensional manifold.
We introduce a neutral metric $\eta: T\mathcal{M}  \times T\mathcal{M}
 \rightarrow \mathbb{R}$ and a bundle map $K:T\mathcal{M} \rightarrow
 T\mathcal{M}$ that satisfies $K^2=1$ and
\begin{align}
  N_K(X,Y) = \frac{1}{4} ([K(X),K(Y)] + [X,Y] - K([K(X),Y] + [X,K(Y)]) = 0,
\label{eq:NijK}
\end{align}
for any $X,Y \in \Gamma (T\mathcal{M})$.
Here, $N_K$ is called the Nijenhuis tensor associated with $K$ and
 \eqref{eq:NijK} is known to be the integrability condition of $K$.
The bundle map $K$ satisfying \eqref{eq:NijK} and $K^2 = 1$ is called
 the para-complex structure.
The set $(\mathcal{M},K,\eta)$ is called a para-Hermitian manifold.

Due to the para-complex structure $K$, the tangent bundle $T\mathcal{M}$
is decomposed into the eigen bundles $L$ and $\tilde{L}$.
They are defined by the following projection operators:
\begin{align}
	P = \frac{1}{2}(1+K),\ \ \tilde{P} = \frac{1}{2}(1- K).
\end{align}
Next, we discuss the integrability of $L$ and $\tilde{L}$.
If the Lie bracket $[\cdot, \cdot]_L$ on $L$ (resp. $\tilde{L}$)
belongs to $L$ (resp. $\tilde{L}$), then it is called involutive.
The involutivity of $L$ ($\tilde{L}$) is rephrased as the vanishing condition of
the tensors $N_P$ ($N_{\tilde{P}}$) defined below:
\begin{align}
N_P(X,Y) = \tilde{P}[P(X),P(Y)],\ \  N_{\tilde{P}}(X,Y) =
 P[\tilde{P}(X),\tilde{P}(Y)], \quad X,Y \in \Gamma (T\mathcal{M}).
\end{align}
By the Frobenius theorem, $L$ ($\tilde{L}$) becomes integrable if and
only if $L$ ($\tilde{L}$) is involutive.
Since $N_K(X,Y) = N_P(X,Y) + N_{\tilde{P}}(X,Y)$,
if $\mathcal{M}$ is a para-Hermitian manifold and $K$ is
integrable, then this means $L$ and $\tilde{L}$ are both integrable.
There is an alternative representation of the Frobenius theorem.
A subbundle in $T\mathcal{M}$ is integrable if and only if a regular
foliation is defined in $\mathcal{M}$.
Therefore, for a para-Hermitian manifold $\mathcal{M}$,
there are foliation structures $\mathcal{F}$ and $\tilde{\mathcal{F}}$ in $\mathcal{M}$ such
that $L = T\mathcal{F}$ and $\tilde{L} = T\mathcal{\tilde{F}}$.
In this case, the local coordinate $x^\mu$ ($\tilde{x}_\mu$) is defined
along a leaf of $\mathcal{F}$ ($\tilde{\mathcal{F}}$).
Namely, a leaf of $\mathcal{F}$ is characterized by $\tilde{x}_\mu =
\text{const}$.
The same is true for $\mathcal{\tilde{F}}$.
Let the basis of $L$ and $\tilde{L}$ be $\partial_\mu$ and $\tilde{\partial}^\mu$.
A vector on $L$ and $\tilde{L}$ is expressed by
$X = X^\mu \partial_\mu \in \Gamma(L)$, $\xi = \xi_\mu
\tilde{\partial}^\mu \in \Gamma(\tilde{L})$.
Therefore, the ordinary Lie brackets $[\cdot,\cdot]_L$ on $L$ and
$[\cdot,\cdot]_{\tilde{L}}$ on $\tilde{L}$ are defined as
\begin{align}
	[X, Y]_L
	&= (X^\mu \partial_\mu Y^\nu - Y^\mu \partial_\mu X^\nu ) \partial_\nu, \notag\\
	[\xi, \eta]_{\tilde{L}}
	&= (\xi_\mu \tilde{\partial}^\mu \eta_\nu - \eta_\mu \tilde{\partial}^\mu \xi_\nu ) \tilde{\partial}^\nu.
  \label{eq:Liebra}
\end{align}
Thus, we can naturally introduce Lie algebroid structures on $L$ and
$\tilde{L}$.

Since $T\mathcal{M} = L \oplus \tilde{L}$, a vector in
$\Gamma(T\mathcal{M})$ is given by
$ e = X + \xi = X^\mu \partial_\mu + \xi_\mu \tilde{\partial}^\mu $.
We call this the doubled vector.
We also introduce the doubled $1$-form $q = q_M \dop x^M = \eta_\mu \dop
 x^\mu + Y^\mu \dop\tilde{x}_\mu$ in $T^*\mathcal{M} \sim L^* \oplus \tilde{L}^*$.
As with the usual tangent bundle, we can define inner product $\langle
 \cdot,\cdot \rangle$ between $T\mathcal{M}$ and $T^*\mathcal{M}$.
Since $\eta$ is a map from
$T\mathcal{M} = L \oplus \tilde{L}$ to
$T^*\mathcal{M} = L^* \oplus \tilde{L}^* $, $\eta$ gives the following isomorphic maps.
 \begin{align}
 	\phi^+ : \tilde{L} \to L^*
 	\qquad \mbox{and} \qquad
 	\phi^- : L \to \tilde{L}^*.
 \end{align}
 These maps imply that the vectors on $\tilde{L}$ is identified with the $1$-form on $L^*$.
 In other words, the basis $\tilde{\partial}^\mu$ on $\tilde{L}$
is identified with the basis $\dop x^{\mu}$ on $L^*$ by $\phi^+$.
 The same is true for vectors on  $L$ and $\tilde{L}^*$ by $\phi^-$.
As a result, the following maps are obtained.
 \begin{align}
 	\Phi^+ : T{\mathcal M} \to L \oplus L^*
 	\qquad \mbox{and} \qquad
 	\Phi^- : T{\mathcal M} \to \tilde{L} \oplus \tilde{L}^*.
   \label{eq:naturaliso}
 \end{align}
In particular, $\Phi^+$ is called the natural isomorphism
\cite{Freidel:2017yuv}.
This provides an explicit relation between the doubled and generalized geometries.
 A doubled vector $X^\mu \partial_\mu + \xi_\mu \tilde{\partial}^\mu $
 in the doubled geometry corresponds to a generalized vector
$X^\mu \partial_\mu + \xi_\mu \dop x^\mu$ in the generalized geometry \cite{Hitchin:2004ut, Gualtieri}.
Finally, we note that when the neutral metric $\eta$ is flat
 \begin{align}
   \eta_{MN}
   = \left(
   \begin{array}{cc}
     0 & 1 \\
     1 & 0
   \end{array}
   \right),
\label{eq:Odd_metric}
 \end{align}
it corresponds to the $O(D,D)$ invariant metric in DFT.

 \subsection{Ante-Courant algebroids on para-Hermitian manifold}

Next, we consider the ante-Courant algebroid on $\mathcal{M}$.
By the general discussion in Section 2,
the doubled structure of an ante-Courant algebroid is compatible with
the {\sf C}-bracket when the anchor satisfies the equation \eqref{eq:anchor_sym}.
We examine this condition in the para-Hermitian manifold $\mathcal{M}$.
Since $E$ and $E^*$ in the general discussion correspond to $L$ and
$\tilde{L}$ in the para-Hermitian manifold, we first write down the
anchor structures in each Lie algebroid.
The anchor $\rho_L: L \to T\mathcal{M}$ on the Lie algebroid $L$ is
expressed as
 \begin{align}
  \rho_L(X)
  &= (\rho_L)^M_{\ \nu}X^\nu\partial_M \notag\\
  \label{eq:anchor_L}
  &= \rho^{\mu}_{\ \nu}X^\nu\partial_\mu + \rho_{\mu\nu}X^\nu \tilde{\partial}^\mu,
\end{align}
where $X \in \Gamma(L)$.
Note that the target of $\rho_L$ is $T\mathcal{M}$.
The adjoint $\rho_L^*$ is defined through the following relation
\begin{align}
  &\langle q ,\rho_L (X) \rangle \notag\\
  &= (\rho^t)_\mu^{\ \nu}\eta_\nu X^\mu + (\rho^t)_{\nu\mu}Y^\mu X^\nu \notag\\
  &= \langle \rho_L^*(q),X \rangle,
\end{align}
where $q = \eta + Y \in \Gamma(T^*\mathcal{M})$ and the symbol $t$ means
transposition of a matrix.
From this expression, we write
$(\rho_L^*)_\mu^{\ N} = ( (\rho^t)_\mu^{\ \nu} ,  (\rho^t)_{\mu\nu})$.
Likewise, the anchor $\rho_{\tilde{L}} : \tilde{L} \to T{\mathcal{M}}$ on
$\tilde{L}$ and its adjoint are expressed as
$
(\rho_{\tilde{L}})^{M\nu}
=
(
\tilde{\rho}^{\mu\nu},
\tilde{\rho}_\mu^{\ \nu}
), \
(\rho_{\tilde{L}}^*)^{\mu N}
=
(
(\tilde{\rho}^t)^{\mu\nu}, (\tilde{\rho}^t)^\mu_{\ \nu}
)
$
.
Therefore, the anchor $\rho_{\rm V} = \rho_L + \rho_{\tilde{L}}$
on $T\mathcal{M} = L\oplus \tilde{L}$ is given by
\begin{align}
  \label{eq:anchor_local}
  (\rho_{\rm V})^M_{\ \ N} =
  \left (
  \begin{array}{cc}
    \rho^\mu_{\ \nu} & \tilde{\rho}^{\mu\nu} \\
    \rho_{\mu\nu} & \tilde{\rho}_\mu^{\ \nu}
  \end{array}
  \right ).
\end{align}
The component expression of
$
\rho_L \rho_{\tilde{L}}^* + \rho_{\tilde{L}} \rho_L^*
$
is
\begin{align}
\bigl(
\rho_L \rho_{\tilde{L}}^*
+
\rho_{\tilde{L}} \rho_L^*
\bigr)^{MN}
  = \left(
  \begin{array}{cc}
    \rho^\mu_{\ \sigma}(\tilde{\rho}^t)^{\sigma\nu} + (\tilde{\rho})^{\mu\sigma}(\rho^t)_\sigma^{\ \nu} & \rho^\mu_{\ \sigma}(\tilde{\rho}^t)^\sigma_{\ \nu} + (\tilde{\rho})^{\mu\sigma}(\rho^t)_{\sigma\nu} \\
    \rho_{\mu\sigma}(\tilde{\rho}^t)^{\sigma\nu} + (\tilde{\rho})_\mu^{\ \sigma}(\rho^t)_\sigma^{\ \nu} & \rho_{\mu\sigma}(\tilde{\rho}^t)^\sigma_{\ \nu} + (\tilde{\rho})_\mu^{\ \sigma}(\rho^t)_{\sigma\nu}
  \end{array}
  \right)
  \label{eq:rho_skew}
\end{align}
Since $\dop_0$ is the exterior derivative on $T\mathcal{M}$, this is given by
$\dop_0 f = \partial_M f \dop x^M= \partial_\mu f \dop x^\mu +
\tilde{\partial}^\mu f \dop \tilde{x}_\mu \in \Gamma(T^*\mathcal{M})$
for $f \in C^\infty(\mathcal{M})$
and the condition \eqref{eq:rho_skew} is expressed by
\begin{align}
\bigl(
\rho_L \rho_{\tilde{L}}^*
+
\rho_{\tilde{L}} \rho_L^*
\bigr)(\dop_0 f) \cdot g
=
\bigl(
\rho_L \rho_{\tilde{L}}^*
+
\rho_{\tilde{L}} \rho_L^*
\bigr)^{MN} \partial_M f \partial_N g
=0.
  \label{eq:C4_local}
\end{align}

Now, we consider a concrete example of $\rho_{\rm V}$ in
\eqref{eq:anchor_local} in the flat para-Hermitian manifold.
The most natural candidate of the anchor is given by the diagonal form
\begin{align}
  (\rho_{\rm V})^M {}_N
  &= \left (
  \begin{array}{cc}
    \rho^\mu_{\ \nu} & 0 \\
    0 & \tilde{\rho}_\mu^{\ \nu}
  \end{array}
  \right ).
  \label{eq:anchor_c}
\end{align}
In particular, the simplest example is
$(\rho_L)^M {}_{\nu} = ( \delta^\mu_{\ \nu}, 0), \,
(\rho_{\tilde{L}})_{\mu} {}^N = ( 0 , \delta_\mu^{\ \nu} )$.
In this case, the condition \eqref{eq:C4_local} is given by
\begin{align}
0 =   ( \rho_* \rho^* + \rho \rho_*^* )^{MN} \partial_N f \partial_M g
  &=
\left(
  \partial_\nu f , \tilde{\partial}^\nu f
\right)
  \left(
  \begin{array}{cc}
    0 & \rho^\mu_{\ \ \sigma} (\tilde{\rho})^\sigma_{\ \nu} \\
    \tilde{\rho}_\mu^{\ \sigma} (\rho^t)_\sigma^{\ \nu} & 0
  \end{array}
  \right)
  \left(
  \begin{array}{c}
    \partial_\nu g \\
    \tilde{\partial}^\nu g
  \end{array}
  \right) \notag\\
  &= \eta^{MN} \partial_M f \partial_N g.
  \label{eq:sc_func}
\end{align}
Here $\eta^{MN}$ is the $O(D,D)$ invariant metric \eqref{eq:Odd_metric} in DFT.
That is, the condition for an ante-Courant algebroid on
a flat para-Hermitian manifold $\mathcal{M}$ is nothing but the strong constraint
only for functions $f,g$.
Since the strong constraint is given by $\eta^{MN} \partial_M \Psi
\partial_N \Phi = 0$ for any quantities $\Psi, \Phi$ in the para-Hermitian
manifold $\mathcal{M}$, the condition \eqref{eq:sc_func} is a relaxed
version of the constraint.
This is consistent with the result in \cite{Chatzistavrakidis:2018ztm}.

\subsection{Pre-Courant algebroids on para-Hermitian manifold}

Next, we examine the conditions for a pre-Courant algebroid on
$\mathcal{M}$.
As we have clarified in Section 3, the condition is only \eqref{eq:prop3.4_v}.
We write down this condition in the flat para-Hermitian manifold.
In particular, when $\rho_{\rm V}$ is given by \eqref{eq:anchor_c}
and $\rho^{\mu} {}_{\nu} = \delta^{\mu} {}_{\nu}$, $\tilde{\rho}_{\mu}
{}^{\nu} = \delta_{\mu} {}^{\nu}$,
we have $X = X^\mu \partial_\mu \in \Gamma(L),
 \dop_* f = \tilde{\partial}^\mu f \partial_\mu,
 \dop f = \partial_\mu f \tilde{\partial}^\mu$
\cite{Mori:2019slw}.
Therefore, the first condition in \eqref{eq:prop3.4_v} is found to be
\begin{align}
0 &=  \mathcal{L}_{\dop f}X + [\dop_* f, X]
\notag \\
  &= \partial_\nu f \tilde{\partial}^\nu X^\mu \partial_{\mu}
 + \tilde{\partial}^\nu f \partial_\nu X^\mu \partial_{\mu}
\notag\\
  &= \eta^{MN}\partial_M f \partial_N X^\mu \partial_{\mu}.
\label{eq:sc_func_vec}
\end{align}
The same is true for $\xi\in\Gamma(\tilde{L})$.
The second condition in \eqref{eq:prop3.4_v} is
\begin{align}
  \eta^{MN} \partial_M f \partial_N \xi_\mu \tilde{\partial}^{\mu}= 0.
  \label{eq:sc_func_form}
\end{align}
Since any pre-Courant algebroids are ante-Courant algebroids,
the condition \eqref{eq:sc_func} is also satisfied.
These conditions \eqref{eq:sc_func_vec} and \eqref{eq:sc_func_form}
are nothing but the strong constraint for
$f \in C^\infty(\mathcal{M})$, $X\in\Gamma(L)$ and $\xi \in\Gamma(\tilde{L})$.
This is again a relaxed version of the strong constraint in DFT.

As noted in Section 3.2, \eqref{eq:sc_func_form} is the necessary
condition for a Poisson structure. In this case, we have the following bracket,
\begin{align}
\{g , f \} = \tilde{\partial}^{\mu} g \partial_{\mu} f = -
 \tilde{\partial}^{\mu} f \partial_{\mu} g = - \{f, g\}.
\label{eq:pC_Poisson}
\end{align}
We find that the skew-symmetric nature is guaranteed by the condition
\eqref{eq:sc_func}.
The Jacobiator of the bracket is calculated as
\begin{align}
\mathrm{Jac} (f,g,h) =& \
\tilde{\partial}^{\mu} (\tilde{\partial}^{\nu} f) \partial_{\nu} g
 \partial_{\mu} h
+
\tilde{\partial}^{\nu} f \tilde{\partial}^{\mu} (\partial_{\nu} g)
 \partial_{\mu} h
+
\tilde{\partial}^{\mu} (\tilde{\partial}^{\nu} g) \partial_{\nu} h
 \partial_{\mu} f
\notag \\
& \
+ \tilde{\partial}^{\nu} g \tilde{\partial}^{\mu} (\partial_{\nu} h)
 \partial_{\mu} f
+ \tilde{\partial}^{\mu} (\tilde{\partial}^{\nu} h) \partial_{\nu} f
 \partial_{\mu} g
+ \tilde{\partial}^{\nu} h \tilde{\partial}^{\mu} (\partial_{\nu} f)
 \partial_{\mu}g.
\end{align}
Due to the conditions \eqref{eq:sc_func_vec} and \eqref{eq:sc_func_form}, we
find $\mathrm{Jac} (f,g,h) = 0$ and confirm that \eqref{eq:pC_Poisson}
indeed defines a Poisson structure.

\subsection{Twisted DFT algebroids}

We have been focusing on the doubled structures of the algebroids and
we clarified the conditions for vectors, forms and functions in the
doubled space-time.
One can introduce additional structures known as the twist by background
fluxes in the manifold.
For example, the twist of the standard Courant algebroid defined by
the generalized tangent bundle $TM \oplus T^*M$ over $M$
has been discussed \cite{Courant}.
The background $3$-form $H$ modifies the standard Courant bracket
$[\cdot,\cdot]_{\rm c}$ giving a new bracket $[\cdot,\cdot]_H$:
\begin{align}
  &[X_1 + \xi_1,X_2 + \xi_2]_{\rm c} = [X_1,X_2] + \mathcal{L}_{X_1}\xi_2 - \mathcal{L}_{X_2}\xi_1 + \frac{1}{2}\dop ( \langle \xi_1 X_2\rangle - \langle \xi_2 X_1\rangle ), \notag\\
  &[X_1 + \xi_1,X_2 + \xi_2]_H = [X_1 + \xi_1, X_2 + \xi_2]_{\rm c} +
 \iop_{X_2}\iop_{X_1}H,
\end{align}
where, $X_i\in\Gamma(TM), \xi_i\in\Gamma(T^*M)$.
The twisted bracket $[\cdot,\cdot]_H$ preserves the Courant algebroid
structure when $\dop H =0$ \cite{Severa:2001qm}.
Physically, this $3$-form $H$ corresponds to the $H$-flux that appears in
the NS-NS sector of type II supergravity.
It is known that this $H$-flux is related to the other fluxes $f,Q,R$ in type II
string theory via the T-duality transformations.
This is represented by the following form \cite{Shelton:2005cf}:
\begin{align}
  H_{abc} \xleftrightarrow{T_c} f_{ab}^{\ \ c}
          \xleftrightarrow{T_b} Q_a^{\ bc}
          \xleftrightarrow{T_a} R^{abc}.
\label{eq:T-dual_chain}
\end{align}
The twist of the Courant algebroid with the ${\sf C}$-bracket has been discussed in \cite{Chatzistavrakidis:2018ztm, Blumenhagen:2012pc}.
In this section, we study compatibility conditions for the doubled and
the twisted structures of the other DFT algebroids.

In order to introduce the twist structure, we consider a doubled $(2,1)$-tensor
$F = F_{MN}{}^{L} \dop x^M \otimes \dop x^N \otimes \partial_L$ on a
flat para-Hermitian manifold $\mathcal{M}$\footnote{
We prefer to use a $(2,1)$-tensor rather than a 3-form on $\mathcal{M}$.
We never introduce ``the generalized doubled tangent bundle'' over $\mathcal{M}$.
}.
We then define the twisted {\sf C}-bracket $[\cdot,\cdot]_{\rm F}$ as follows:
\begin{align}
  [e_1,e_2]_{\rm F} = [e_1,e_2]_{\sf C} + \iop_{e_2}\iop_{e_1}F, \quad
e_i \in\Gamma (\mathcal{V}).
\end{align}
Here $\iop_{e_i}:\Gamma (\wedge^{p} \mathcal{V}^*) \to \Gamma
 (\wedge^{p-1}\mathcal{V}^*)$
is the interior product defined by
\begin{align}
  (\iop_{e_i} q) (a_1, \ldots, a_{p-1}) = h (e_i, a_1, \ldots, a_{p-1}),\ \ \ q\in\Gamma(\wedge^{p}\mathcal{V}^*),\ a_1,\cdots,a_{p-1} \in\Gamma(\mathcal{V}).
\end{align}
In the following, we assume that $(\mathcal{V}, [\cdot,\cdot]_{\sf C},
\rho_{\rm V}, \pbra{\cdot,\cdot})$ is a Vaisman algebroid with the
doubled structure discussed in the previous sections.

\paragraph{Twisted Vaisman algebroid}
 We consider $(\mathcal{V}, [\cdot,\cdot]_{\rm F}, \rho_{\rm V},
 \pbra{\cdot,\cdot})$.
and examine Axiom V1 (C3) and Axiom V2 (C5).
Axiom V1, the equation \eqref{eq:V1}, gives the Leibniz rule for the bracket.
Expanding the left-hand side of the equation \eqref{eq:V1},
we find
\begin{align}
  [e_1,f e_2]_{\rm F}
  & = [e_1,f e_2]_{\sf C} + \iop_{e_2}\iop_{e_1}F \notag\\
  & = f[e_1,e_2]_{\sf C} + f\iop_{e_2}\iop_{e_1}F + (\rho_{\rm V}(e_1)f)e_2 - \pbra{e_1,e_2}\mathcal{D} f \notag\\
  & = f[e_1,e_2]_{\rm F} + (\rho_{\rm V}(e_1)f)e_2 - \pbra{e_1,e_2}\mathcal{D} f,
\end{align}
where we have used the property of the {\sf C}-bracket.
Therefore, $(\mathcal{V}, [\cdot,\cdot]_{\rm F}, \rho_{\rm V},
\pbra{\cdot,\cdot})$ satisfies Axiom V1 automatically.

Axiom V2, the equation \eqref{eq:V2}, is a compatibility condition between $\pbra{\cdot,\cdot}$ and $\rho_{\rm V}$.
The right-hand side of the equation \eqref{eq:V2} is evaluated as
\begin{align}
  & \pbra{[e_1,e_2]_{\rm F} + \mathcal D\pbra{e_1,e_2} , e_3} + \pbra{e_2 , [e_1,e_3]_{\rm F} + \mathcal D\pbra{e_1,e_3}} \notag\\
  &= \pbra{[e_1,e_2]_{\rm V} + \mathcal D\pbra{e_1,e_2} , e_3} + \pbra{e_2 , [e_1,e_3]_{\rm V} + \mathcal D\pbra{e_1,e_3}} \notag\\
  &\quad + \pbra{\iop_{e_2}\iop_{e_1}F, e_3} + \pbra{e_2, \iop_{e_3}\iop_{e_1}F }.
\end{align}
The first and the second terms on the right-hand side become
$\rho_{\rm V}(e_1)\pbra{e_2,e_3}$.
Thus, Axiom V2 is satisfied when the third and the fourth terms vanish:
\begin{align}
  \pbra{\iop_{e_2}\iop_{e_1}F, e_3} + \pbra{e_2, \iop_{e_3}\iop_{e_1}F }=0\ \ \ \ {}^{\forall}e_1,e_2,e_3\in\Gamma(\mathcal{V}).
  \label{eq:ante_F_cond}
\end{align}
Since the basis of $\Gamma(\mathcal{V})$ is
$\partial_M = (\partial_\mu, \tilde{\partial}^\mu)$,
$\iop_{e_2}\iop_{e_1}F$ is given by
\begin{align}
  \iop_{e_2}\iop_{e_1}F
  &= (e_1)^M (e_2)^N F_{MN}^{\ \ \ \ l} \partial_l + (e_1)^M (e_2)^N F_{MNl} \tilde{\partial}^l.
  \label{eq:decompose_F}
\end{align}
Then, we can denote $\pbra{\cdot,\cdot}$ as
\begin{align}
  \pbra{e_1,e_2}
  = \frac{1}{2}\bigl(\langle \xi_1, X_2 \rangle + \langle \xi_2, X_1 \rangle \bigl)
  = \frac{1}{2} \eta_{MN} (e_1)^M (e_2)^N.
\end{align}
Here, $X_i\in\Gamma(L),\xi_i\in\Gamma(\tilde{L}^*)$ and $e_i = X_i + \xi_i$.
Therefore, the condition \eqref{eq:ante_F_cond} becomes
\begin{align}
  0
  &=\pbra{\iop_{e_2}\iop_{e_1}F, e_3} + \pbra{e_2, \iop_{e_3}\iop_{e_1}F} \notag\\
  &= \frac{1}{2} ( \eta_{KL} F_{MN}^{\ \ \ \ K} + \eta_{NK} F_{ML}^{\ \ \ \ K} )(e_1)^M (e_2)^N (e_3)^L.
  \label{C5:compornent}
\end{align}
Namely,
\begin{align}
  F_{MNL} + F_{MLN} = 0.
  \label{eq:C3cond_local}
\end{align}
Here the doubled indices are raised and lowered by the $O(D,D)$
invariant metric $\eta_{MN}$ and its inverse $\eta^{MN}$.
In summary,
$(\mathcal{V}, [\cdot,\cdot]_{\rm F}, \rho_{\rm V},
 \pbra{\cdot,\cdot})$
becomes a twisted Vaisman algebroid only when the condition
\eqref{eq:C3cond_local} is satisfied.
This means that the tensor $F_{MNL}$ is anti-symmetric with respect to
the latter two indices.
We note that the doubled tensor $F_{MN} {}^K$ is decomposed as
$F_{MN} {}^K = (H_{\mu \nu \rho}, f_{\mu \nu} {}^{\rho}, Q_{\mu} {}^{\nu
\rho}, R^{\mu \nu \rho})$ involving all the fluxes in \eqref{eq:T-dual_chain}.

\paragraph{Twisted ante-Courant algebroid}
Next we discuss a twisted ante-Courant algebroid with the doubled
structure.
We assume that $(\mathcal{V}, [\cdot,\cdot]_{\sf C}, \rho_{\rm V},
\pbra{\cdot,\cdot})$ is an ante-Courant algebroid and
look for conditions that $(\mathcal{V}, [\cdot,\cdot]_{\rm F}, \rho_{\rm
V}, \pbra{\cdot,\cdot})$ becomes also an ante-Courant algebroid.
Since any ante-Courant algebroids are Vaisman algebroids, the tensor $F$
should satisfy the condition \eqref{eq:C3cond_local}.
As we have discussed, the condition for the ante-Courant algebroid is
\eqref{eq:anchor_sym}.
However, this is the condition for the anchor map which is
irrelevant to the bracket structure.
Therefore, we need no extra conditions for $F$.

\paragraph{Twisted pre-Courant algebroid}
Next, we discuss a twisted pre-Courant algebroid.
Again we assume that $(\mathcal{V}, [\cdot,\cdot]_{\sf C}, \rho_{\rm V},
\pbra{\cdot,\cdot})$ is a pre-Courant algebroid.
We write down the conditions that
$(\mathcal{V}, [\cdot,\cdot]_{\rm F}, \rho_{\rm V},
\pbra{\cdot,\cdot})$ becomes a pre-Courant algebroid.
In addition to the condition \eqref{eq:C3cond_local}, we need
Axiom C2, namely, the homomorphism of $\rho_{\rm V}$ \eqref{eq:C2}.
The left-hand side of the equation in \eqref{eq:C2} is evaluated as
\begin{align}
  \rho_{\rm V}([e_1,e_2]_{\rm F})
  &= \rho_{\rm V}([e_1,e_2]_{\rm V} + \iop_{e_2}\iop_{e_1}F) \notag\\
  &= [\rho_{\rm V}(e_1),\rho_{\rm V}(e_2)] + \rho_{\rm V}\bigl(\iop_{e_2}\iop_{e_1}F\bigr).
\end{align}
Thus, the condition is
\begin{align}
  \rho_{\rm V}\bigl( \iop_{e_2}\iop_{e_1}F\bigr)
  &= (e_1)^M(e_2)^N (\rho_{\rm V})^L_{\ \ K} F_{MN}^{\ \ \ \ K}
 \partial_L = 0.
\end{align}
In component, we have,
\begin{align}
  (\rho_{\rm V})^L_{\ K} F_{MN}^{\ \ \ \ K} = 0.
  \label{eq:twist_C2}
\end{align}
Then for the non-zero tensor $F$, the anchor should satisfy
\begin{align}
\det \rho_{\rm V} = 0.
\label{eq:det_rho}
\end{align}
Therefore, we should add the condition not only for the tensor
 \eqref{eq:C3cond_local} but also for the anchor \eqref{eq:det_rho}
to obtain a twisted pre-Courant algebroid
$(\mathcal{V}, [\cdot,\cdot]_{\rm F}, \rho_{\rm V}, \pbra{\cdot,\cdot})$.
In particular, when \eqref{eq:anchor_c} is adapted for $\rho_{\rm
V}$, either $\rho_{L}$ or $\rho_{\tilde{L}}$ must be zero.

\paragraph{Twisted Courant algebroid}
Finally, we consider a twisted Courant algebroid.
We assume that $(\mathcal{V}, [\cdot,\cdot]_{\sf C}, \rho_{\rm V},
\pbra{\cdot,\cdot})$ is a Courant algebroid.
We calculate the Jacobiator \eqref{eq:C1} for $[e_1,e_2]_{\rm F}$ and confirm Axiom C1.
The result is
\begin{align}
  [[e_1,e_2]_{\rm F}, e_3]_{\rm F} + {\rm c.p.}
  &= \mathcal{D} T_{\rm F}(e_1,e_2,e_3) \notag\\
  &\quad - \frac{1}{3} \mathcal{D} \bigl( \pbra{\iop_{e_2}\iop_{e_1}F, e_3} + {\rm c.p.}\bigr) \notag\\
  &\quad + \bigl( \iop_{e_3}\iop_{([e_1,e_2]_{\rm V})}F + [\iop_{e_2}\iop_{e_1}F, e_3]_{\rm V} + \iop_{e_3}\iop_{(\iop_{e_2}\iop_{e_1}F)}F + {\rm c.p.} \bigr).
  \label{eq:Jac_F}
\end{align}
Here $T_{\rm F}$ is defined by
\begin{align}
  T_{\rm F}(e_1,e_2,e_3)
  &= T(e_1,e_2,e_3) + \frac{1}{3}\bigl( \pbra{\iop_{e_2}\iop_{e_1}F, e_3} + {\rm c.p.}\bigr).
  \label{eq:TandT_F}
\end{align}
Then the condition for Axiom C1 is
\begin{align}
  & - \frac{1}{3} \mathcal{D} \bigl( \pbra{\iop_{e_2}\iop_{e_1}F, e_3} + {\rm c.p.}\bigr) \notag\\
  &\quad + \bigl( \iop_{e_3}\iop_{([e_1,e_2]_{\rm V})}F + [\iop_{e_2}\iop_{e_1}F, e_3]_{\rm V} + \iop_{e_3}\iop_{(\iop_{e_2}\iop_{e_1}F)}F + {\rm c.p.} \bigr) = 0.
  \label{condition:Jac_F}
\end{align}
Since $(\mathcal{V}, [\cdot,\cdot]_{\rm F}, \rho_{\rm V},
\pbra{\cdot,\cdot})$ should be a pre-Courant algebroid,
the anchor satisfies the condition $\det \rho_{\rm V} = 0$.
To solve this condition, we assume that the anchor is given by the
diagonal form \eqref{eq:anchor_c}.
Then one of the anchors $\rho$ or $\tilde{\rho}$ must be zero.
In the following, we select a frame where $\tilde{\rho} = 0$ and
$\rho$ is an identity matrix.
When $\rho_{\tilde{L}} = 0$, this implies $\dop_* f=0$
and $\mathcal{L}_{\xi_i} = 0$ \cite{Mackenzie}.
Since we assumed that $(\mathcal{V}, [\cdot,\cdot]_{\sf C}, \rho_{\rm
V}, \pbra{\cdot,\cdot})$ becomes a Courant algebroid,
the derivation condition \eqref{eq:derivation} must hold.
This is represented by the local coordinate as \cite{Mori:2019slw}
\begin{align}
	0
	=\tilde{\partial}^\rho A^\mu \partial_\rho B^\nu
		+ \tilde{\partial}^\rho B^\nu \partial_\rho A^\mu
	= \eta^{KL} \partial_K A^\mu \partial_L B^\nu, \ \ A,B\in\Gamma(L).
  \label{eq:sc_compornent}
\end{align}
Furthermore, the ``relaxed'' version of the strong constraints
\eqref{eq:sc_func}, \eqref{eq:sc_func_vec} and \eqref{eq:sc_func_form}
are satisfied as discussed before.
The most natural solution to these conditions is obtained by setting
$\tilde{\partial}^\mu \Psi = 0$ for any quantities $\Psi$ in $\mathcal{M}$.
Then the Lie bracket $[\cdot,\cdot]_*$ given in \eqref{eq:Liebra} becomes zero.

With these conditions at hand, we evaluate \eqref{condition:Jac_F}.
The first term in the left-hand side of \eqref{condition:Jac_F} is
expressed as
\begin{align}
  \mathcal{D} \pbra{\iop_{e_2}\iop_{e_1}F, e_3}
  &= \frac{1}{2} \mathcal{D} (\iop_{q_3}\iop_{e_2}\iop_{e_1} F).
\end{align}
Here we have introduced the interior product
$\iop_{q_i} : \Gamma (\wedge^{p} \mathcal{V}) \to \Gamma (\wedge^{p-1}\mathcal{V})$
by a doubled 1-form $q_i$ which acts on a doubled $k$-vector.
Therefore,
\begin{align}
  - \frac{1}{3} \mathcal{D} (\pbra{\iop_{e_2}\iop_{e_1}F, e_3} + {\rm c.p.})
= - \frac{1}{3}
\dop (\iop_{q_3}\iop_{e_2}\iop_{e_1} F + {\rm c.p.}),
  \label{eq:condition:Jac_F1}
\end{align}
where we have used the fact $\mathcal{D} = \dop + \dop_*$.
Likewise, the second term in the left-hand side in \eqref{condition:Jac_F} becomes
\begin{align}
\iop_{e_3}\iop_{[e_1,e_2]_{\sf C}}F + {\rm c.p.}
=
\iop_{e_3}\iop_{( [X_1, X_2]_L + \mathcal{L}_{X_1} \xi_2
 - \mathcal{L}_{X_2} \xi_1 + \dop \mbra{e_1,e_2})}F
+ {\rm c.p.}
 \label{eq:Jac_F2}
\end{align}
The third term in the left-hand side of \eqref{condition:Jac_F} is
\begin{align}
[\iop_{e_2}\iop_{e_1}F, e_3]_{\rm V} + {\rm c.p.}
=& \
[ \iop_{e_2}\iop_{e_1}F^\mu \partial_\mu, X_3] +
\mathcal{L}_{( \iop_{e_2}\iop_{e_1}F^\mu \partial_\mu )} \xi_3
\notag \\
& \
 - \mathcal{L}_{X_3} (\iop_{e_2}\iop_{e_1}F_\mu \tilde{\partial}^\mu )
  + \dop \mbra{ \iop_{e_2}\iop_{e_1}F , e_3 } + {\rm c.p.}
\end{align}
Here we have used the following notations,
\begin{align}
  \iop_{e_2}\iop_{e_1}F^l \partial_l = (e_1)^M (e_2)^N F_{MN}^{\ \ \ \ l} \partial_l, \notag\\
  \iop_{e_2}\iop_{e_1}F_l \tilde{\partial}^l = (e_1)^M (e_2)^N F_{MNl} \tilde{\partial}^l.
\end{align}
Then the condition \eqref{condition:Jac_F} is found to be
\begin{align}
&
- \frac{1}{3} \dop (\iop_{q_3}\iop_{e_2}\iop_{e_1} F)
+ \iop_{e_3}\iop_{( [X_1, X_2]_L + \mathcal{L}_{X_1} \xi_2 -
 \mathcal{L}_{X_2} \xi_1 + \dop \mbra{e_1,e_2})}F
+ [ \iop_{e_2}\iop_{e_1}F^\mu \partial_\mu, X_3]
\notag\\
&
+ \mathcal{L}_{(
\iop_{e_2}\iop_{e_1}F^\mu \partial_\mu )} \xi_3
- \mathcal{L}_{X_3} (\iop_{e_2}\iop_{e_1}F_\mu \tilde{\partial}^\mu )
+ \dop \mbra{ \iop_{e_2}\iop_{e_1}F , e_3 }
+ \iop_{e_3}\iop_{(\iop_{e_2}\iop_{e_1}F)} F
 + {\rm c.p.}
\notag\\
&
= 0.
  \label{condition:Jac_F_decomp}
\end{align}
If $F$ is a totally anti-symmetric tensor
$H_{\mu\nu\rho} \tilde{\partial}^\mu \wedge \tilde{\partial}^\nu \wedge
\tilde{\partial}^\rho$, one can show that
the left-hand side of \eqref{condition:Jac_F_decomp} becomes
$\iop_{X_3}\iop_{X_2}\iop_{X_1}\dop H$.
Therefore when $\dop H =0$, Axiom C1 holds and $(\mathcal{V}, [\cdot,\cdot]_{\rm F}, \rho_{\rm V},
\pbra{\cdot,\cdot})$ becomes the $H$-twisted standard Courant algebroid
known in the literature.
When we consider the condition $\tilde{\partial}^\mu \Phi = 0$,
then as described in Section 4.1, this means that we restrict the doubled
space to a leaf in the foliation $\mathcal{F}$ of $\mathcal{M}$.
This leaf is interpreted as the physical space-time and $\dop H = 0$
is nothing but the Bianchi identity for the field strength of the
NS-NS $B$-field in type II supergravity.

On the other hand, when we consider an alternative frame
$\rho_{L}= 0$, $\partial_{\mu} \Psi = 0$,
we have $\dop f = 0$, $\mathcal{L}_{X_i} = 0$ and $[\cdot, \cdot]_L = 0$.
In this case,
$F = R^{\mu\nu\rho} \partial_\mu \wedge \partial_\nu \wedge
\partial_\rho$ is allowed and $\dop_* R =0$ appears as a condition.
The other possibilities including
$F = f_{\mu\nu}^{\ \ \rho} \tilde{\partial}^\mu \wedge
\tilde{\partial}^\nu \wedge \partial_\rho$
and
$F = Q_{\mu}^{\ \nu\rho} \tilde{\partial}^\mu \wedge \partial_\nu \wedge
\partial_\rho$, would be allowed for general $\rho_{\rm V}$
\eqref{eq:anchor_local}.
In particular, the role of the non-diagonal component $\tilde{\rho}^{\mu\nu}$ in
\eqref{eq:anchor_local} and the tensors $f_{\mu \nu} {}^{\rho}$, $Q_{\mu} {}^{\nu \rho}$
is discussed in \cite{Chatzistavrakidis:2018ztm}.

\section{From algebroids to algebras on group manifolds} \label{sect:group}

In this section, we make a brief comment on the doubled structures
discussed in this paper and those in group manifolds.
The notion of the ``double'' has been originally proposed in the
context of Hopf algebras by Drinfel'd \cite{Drinfeld:1986}.
A classical limit of this operation is implemented in Lie algebras
\cite{Kosmann-Schwarzbach3, Marle}.
A well known fact is that a Lie algebra is defined by the left invariant
vectors at the unit element of a group manifold.
On the other hand, DFT on group manifolds has recently been considered \cite{Blumenhagen:2014gva}.
In this setup, the manifest T-duality of DFT is generalized to the so-called
Poisson-Lie T-duality \cite{Klimcik:1995ux}.
An essential feature of the Poisson-Lie T-duality lies in the structure
of the Drinfel'd double of the underlying group manifold.
Indeed, the Lie algebras associated with the abovementioned group is given by
the Drinfel'd double.
We note that it is possible to introduce the para-Hermitian nature
even for group manifolds \cite{Marotta:2019eqc}.
It is therefore natural to consider the relation between the doubled
structure of the algebroids discussed here and the Lie algebras of the group
manifold.
The left invariant vector fields that define the Lie algebra of the
group manifold are essentially given by a point on the group, namely,
the unit element.
Therefore, in order to find the associated Lie algebras from the
algebroids, we consider only the unit element on the group manifold and
restrict the vector space to the one for the left invariant vectors.
This procedure is achieved by setting $\rho_{\rm V} = 0$.
Then we have $\rho_{E}^{*}\dop_0 = \dop =0, \rho^{*}_{E^{*}}\dop_0 = \dop_* = 0$, $\mathcal{D}=0$.
Under these conditions, Axioms C1 - C5 are rewritten as
\begin{description}
  \item[Axiom C1] ${\rm Jac}(e_1,e_2,e_3)=0$. The {\sf C}-bracket becomes a Lie bracket.
  \item[Axiom C2] This becomes trivial by $\rho_{\rm V} =0$.
  \item[Axiom C3] $[e_1,fe_2]_{\sf C} = f[e_1,e_2]_{\sf C}$ for any $f \in
	     C^\infty(M)$. This shows the bilinearity of the Lie bracket.
  \item[Axiom C4] This becomes trivial by $\mathcal{D}=0$.
  \item[Axiom C5] $ \pbra{ [e_1,e_2]_{\sf C} , e_3 } + \pbra{ e_1 , [e_1,e_3]_{\sf C} }
	     = 0$.
\end{description}
The last one is the compatibility condition between $[e_1,e_2]_{\sf C}$
and $\pbra{\cdot, \cdot}$.
In general,
a Lie algebra that has the compatible bilinear form
$\pbra{\cdot,\cdot}$ is called a quadratic Lie algebra.
It is also known that quadratic Lie algebras are infinitesimal algebras of Poisson-Lie groups.
Therefore, we can see that a quadratic Lie algebra
$(\mathcal{V}, [\cdot,\cdot]_{\sf C}, \pbra{\cdot,\cdot})$ is obtained
by the Courant algebroid made by the Drinfel'd double on the group manifold.

Now let us consider the Vaisman algebroid.
As we have shown, the Jacobi identity is broken by the following
quantities $J_1,J_2$:
\begin{align}
  J_1
  &= \iop_{X_3}
\Bigl(
{\dop}[\xi_1,\xi_2]_{E^*} - \mathcal{L}_{\xi_1}{\dop}\xi_2 +
 \mathcal{L}_{\xi_2}{\dop}\xi_1
\Bigr)
+
\iop_{\xi_3}
\Bigl(
{\dop}_*[X_1,X_2]_E - \mathcal{L}_{X_1}{\dop}_*X_2 +
 \mathcal{L}_{X_2}{\dop}_*X_1
\Bigr),
\notag\\
  J_2
  &=
\Bigl(
\mathcal{L}_{{\dop}_* \mbra{e_1,e_2}}\xi_3 + [{\dop}\mbra{e_1,e_2},\xi_3]_{E^*}
\Bigr)
 -
\Bigl(
\mathcal{L}_{{\dop} \mbra{e_1,e_2}}X_3 + [{\dop}_*\mbra{e_1,e_2},X_3]_E
\Bigr).
\end{align}
When we set $\rho_{\rm V}=0, \dop=0, \dop_*=0$, then it is obvious that $J_1=J_2=0$.
Thus, even if we consider the algebroid where Axiom C1 does not hold,
going to the algebra, we end up with ${\rm Jac}(e_1,e_2,e_3)=0$ and the
bracket becomes a Lie bracket.
Therefore, all the algebroids with doubled structure discussed in this
paper reduce to the quadratic Lie algebras at the unit of the group manifold.
Since the doubled structure of algebroids is essentially irrelevant to
the group action, the DFT algebroids discussed in this paper are
non-group generalizations of the quadratic Lie algebras defined
by the Drinfel'd double.
This would be a key property for the further generalizations of the
Poisson-Lie T-duality.

\section{Conclusion and discussions} \label{sect:conclusion}
We studied doubled aspects of algebroid structures
appearing in double field theory (DFT).
The gauge symmetry is governed by a skew-symmetric bracket called the
{\sf C}-bracket.
The {\sf C}-bracket unifies the two characteristic symmetries in string theory, the
diffeomorphism and the $U(1)$ gauge symmetry of the NSNS
$B$-field, in an $O(D,D)$ T-duality covariant way.
The {\sf C}-bracket suggests doubled structures as a mathematical nature
of DFT.
Indeed, one of the notable features of algebroids equipped with the {\sf
C}-bracket is its doubled nature that evokes the Drinfel'd double for Lie bialgebras.
Among other things, the Vaisman algebroid that appears most
naturally on the tangent bundle of the para-Hermitian
manifold consists of a pair of Lie algebroids on the doubled foliations \cite{Mori:2019slw}.
This is similar to the Drinfel'd double discussed in the Courant algebroid \cite{LiWeXu}.
It has been pointed out that the strong constraint of
DFT, which is the most important condition for the theory to be a
consistent physical theory, is nothing but the necessary condition
that the Lie algebroid pairs become a Lie bialgebroid.
After solving the strong constraint, the Vaisman algebroid reduces to
the Courant algebroid in the physical space-time.
It has also been shown that once the strong constraint is imposed on any
quantities on the para-Hermitian manifold, the gauge algebra in DFT
closes.
In this sense, the consistency condition for the Lie algebroid pairs
is a geometrical realization of the strong constraint in DFT.

In this paper, we first classified possible algebroid structures allowed by the independent axioms among C1-C5.
Other than the Courant and the Vaisman algebroids, we showed that there are additional
algebroid structures defined by parts of Axioms C1-C5.
We showed that there are four distinct algebroids, which we call
the ante-Courant, the pre-Courant, the
Jacobi Vaisman and the Jacobi pre-Courant algebroids.

We then introduced the doubled structure and the {\sf C}-bracket and tried to construct these algebroids explicitly.
Similar to the Vaisman and the Courant algebroids, we found that the
ante- and the pre-Courant algebroids are built out of two dual Lie
algebroids with appropriate conditions.
We also showed that the conditions for the pre-Courant algebroid is necessary for
non-trivial Poisson structures.
On the other hand, we found that the Jacobi Vaisman and the Jacobi ante-Courant algebroids
are inconsistent with the doubled structure and the {\sf C}-bracket.
We nevertheless stress that they are allowed axiomatically.
Although they are not compatible with the doubled structure and the {\sf C}-bracket,
it would be interesting to explore explicit examples of these algebroids based on a consistent bracket.

In the latter part of this paper, we exhibit the concrete examples of
the DFT algebroids in the doubled space-time.
We consider the $2D$-dimensional flat para-Hermitian manifold as a doubled space-time.
We found that the consistency conditions of the doubled structures
for these algebroids are just the relaxed versions of the strong constraint in DFT.
We also studied the consistency conditions for the twisted DFT algebroids.
The twist is introduced by a $(2,1)$-tensor in the para-Hermitian
manifold. We clarify that the tensor should satisfy appropriate
conditions related by the anchors in the Lie algebroid pairs of the
doubled structure.
We showed that the (relaxed versions of) the strong constraint implies the
induced Poisson structure in general.
This means that the Poisson structure is closely related to the doubled nature of algebroids.
Even though, this becomes trivial in the flat para-Hermitian manifold, it still provides non-trivial structures
in the general curved para-Hermitian manifold.

We finally comment on the doubled structures in group manifolds.
The doubled structure discussed in the algebroids is taken over to those for the Lie algebras in the group manifolds.
The nature of the strong constraint, the double, and the Poisson structure are all necessary ingredients for the Poisson-Lie T-duality.
These structures are important when one considers generalizations of T-duality in general curved doubled space.
The doubled structure would also be important to study the global aspects of the doubled space-time \cite{Ikeda:2020lxz}.
It would be interesting to study further this direction.
We will come back to these issues in future studies.

\subsection*{Acknowledgments}
The authors would like to thank K.~Shiozawa for useful discussions.
The work of H.~M. is supported in part by the Sasakawa
Scientific Research Grant (No. 2020-2009) from the Japan Science Society.
The work of S.~S. is supported in part by Grant-in-Aid for Scientific Research, JSPS KAKENHI Grant Number JP20K03952.
\\
\\
The data that supports the findings of this study are available within the article.



\begin{thebibliography}{0}

\bibitem{Siegel:1993xq}
  W.~Siegel,
  ``Two vierbein formalism for string inspired axionic gravity,''
  Phys.\ Rev.\ D {\bf 47} (1993) 5453
  doi:10.1103/PhysRevD.47.5453
  [hep-th/9302036],
  ``Superspace duality in low-energy superstrings,''
  Phys.\ Rev.\ D {\bf 48} (1993) 2826
  doi:10.1103/PhysRevD.48.2826
  [hep-th/9305073],
  ``Manifest duality in low-energy superstrings,''
  hep-th/9308133.

\bibitem{Hull:2009mi}
  C.~Hull and B.~Zwiebach,
  ``Double Field Theory,''
  JHEP {\bf 0909} (2009) 099
  doi:10.1088/1126-6708/2009/09/099
  [arXiv:0904.4664 [hep-th]].

 \bibitem{Vaisman:2012ke}
   I.~Vaisman,
   ``On the geometry of double field theory,''
   J.\ Math.\ Phys.\  {\bf 53} (2012) 033509
   doi:10.1063/1.3694739
   [arXiv:1203.0836 [math.DG]].

  \bibitem{Vaisman:2012px}
    I.~Vaisman,
    ``Towards a double field theory on para-Hermitian manifolds,''
    J.\ Math.\ Phys.\  {\bf 54} (2013) 123507
    doi:10.1063/1.4848777
    [arXiv:1209.0152 [math.DG]].

  \bibitem{Hohm:2010jy}
  O.~Hohm, C.~Hull and B.~Zwiebach,
  ``Background independent action for double field theory,''
  JHEP \textbf{07} (2010), 016
  doi:10.1007/JHEP07(2010)016
  [arXiv:1003.5027 [hep-th]].

  \bibitem{LiWeXu}
    Z-J.~Liu, A.~Weinstein, P.~Xu,
    ``Manin Triples for Lie Bialgebroids,''
    J. \ Differential Geom. {\bf 45} (1997) 547 [dg-ga/9508013].

  \bibitem{Courant}
    T.~J.~Courant,
    ``Dirac Manifolds,''
    Trans. Amer. Math. Soc. {\bf 319} (1990) 631.

  \bibitem{Svoboda:2018rci}
    D.~Svoboda,
    ``Algebroid Structures on Para-Hermitian Manifolds,''
    J. Math. Phys. \textbf{59} (2018) no.12, 122302
    doi:10.1063/1.5040263
    [arXiv:1802.08180 [math.DG]].

  \bibitem{Carow-Watamura:2020xij}
  U.~Carow-Watamura, K.~Miura, S.~Watamura and T.~Yano,
   ``Metric algebroid and Dirac generating operator in Double Field Theory,''
  [arXiv:2005.04658 [hep-th]].

  \bibitem{Chatzistavrakidis:2018ztm}
    A.~Chatzistavrakidis, L.~Jonke, F.~S.~Khoo and R.~J.~Szabo,
    ``Double Field Theory and Membrane Sigma-Models,''
    JHEP {\bf 1807} (2018) 015
    doi:10.1007/JHEP07(2018)015
    [arXiv:1802.07003 [hep-th]].

  \bibitem{Hull:2009zb}
    C.~Hull and B.~Zwiebach,
    ``The Gauge algebra of double field theory and Courant brackets,''
    JHEP {\bf 0909} (2009) 090
    doi:10.1088/1126-6708/2009/09/090
    [arXiv:0908.1792 [hep-th]].

  \bibitem{Mori:2019slw}
    H.~Mori, S.~Sasaki and K.~Shiozawa,
    ``Doubled Aspects of Vaisman Algebroid and Gauge Symmetry in Double Field Theory,''
    J. Math. Phys. \textbf{61} (2020) no.1, 013505
    doi:10.1063/1.5108783
    [arXiv:1901.04777 [hep-th]].

  \bibitem{Klimcik:1995jn}
    C.~Klimcik,
    ``Poisson-Lie T duality,''
    Nucl.\ Phys.\ Proc.\ Suppl.\  {\bf 46} (1996) 116
    doi:10.1016/0920-5632(96)00013-8
    [hep-th/9509095].

  \bibitem{Klimcik:1995ux}
    C.~Klimcik and P.~Severa,
    ``Dual nonAbelian duality and the Drinfeld double,''
    Phys.\ Lett.\ B {\bf 351} (1995) 455
    doi:10.1016/0370-2693(95)00451-P
    [hep-th/9502122].

  \bibitem{VonUnge:2002xjf}
    R.~Von Unge,
    ``Poisson Lie T plurality,''
    JHEP {\bf 0207} (2002) 014
    doi:10.1088/1126-6708/2002/07/014
    [hep-th/0205245].

  \bibitem{Sakatani:2019zrs}
    Y.~Sakatani,
    ``U-duality extension of Drinfel'd double,''
    PTEP {\bf 2020} (2020) no.2,  023B08
    doi:10.1093/ptep/ptz172
    [arXiv:1911.06320 [hep-th]].

  \bibitem{Malek:2019xrf}
  E.~Malek and D.~C.~Thompson,
  ``Poisson-Lie U-duality in Exceptional Field Theory,''
  JHEP \textbf{04} (2020), 058
  doi:10.1007/JHEP04(2020)058
  [arXiv:1911.07833 [hep-th]].

  \bibitem{Blair:2020ndg}
  C.~D.~A.~Blair, D.~C.~Thompson and S.~Zhidkova,
  ``Exploring Exceptional Drinfeld Geometries,''
  [arXiv:2006.12452 [hep-th]].

  \bibitem{Malek:2020hpo}
  E.~Malek, Y.~Sakatani and D.~C.~Thompson,
  ``E$_{6(6)}$ Exceptional Drinfel'd Algebras,''
  [arXiv:2007.08510 [hep-th]].

  \bibitem{Hassler:2017yza}
	F.~Hassler,
         ``Poisson-Lie T-Duality in Double Field Theory,''
	Phys. Lett. B \textbf{807} (2020), 135455
	doi:10.1016/j.physletb.2020.135455
	[arXiv:1707.08624 [hep-th]].

	\bibitem{Demulder:2018lmj}
	S.~Demulder, F.~Hassler and D.~C.~Thompson,
	``Doubled aspects of generalised dualities and integrable deformations,''
	JHEP \textbf{02} (2019) 189
	doi:10.1007/JHEP02(2019)189
	[arXiv:1810.11446 [hep-th]].

  \bibitem{Sakatani:2019jgu}
    Y.~Sakatani,
    ``Type II DFT solutions from Poisson-Lie T-duality/plurality,''
    PTEP (2019) 073B04
    doi:10.1093/ptep/ptz071
    [arXiv:1903.12175 [hep-th]].

  \bibitem{Hlavaty:2019pze}
  L.~Hlavat\'y and I.~Petr,
  ``Poisson-Lie plurals of Bianchi cosmologies and Generalized Supergravity Equations,''
  JHEP \textbf{04} (2020), 068
  doi:10.1007/JHEP04(2020)068
  [arXiv:1910.08436 [hep-th]].

  \bibitem{Musaev:2020bwm}
  E.~T.~Musaev,
  ``On non-abelian U-duality of 11D backgrounds,''
  [arXiv:2007.01213 [hep-th]].

  \bibitem{Delduc:2013qra}
    F.~Delduc, M.~Magro and B.~Vicedo,
    ``An integrable deformation of the $AdS_5 \times S^5$ superstring action,''
    Phys.\ Rev.\ Lett.\  {\bf 112} (2014) no.5,  051601
    doi:10.1103/PhysRevLett.112.051601
    [arXiv:1309.5850 [hep-th]].

  \bibitem{Mackenzie}
    K.~C.~H.~Mackenzie and P.~Xu,
    ``Lie bialgebroids and Poisson groupoids,''
    Duke Math. \ J. {\bf 73} (1994) 415.

  \bibitem{Uchino:2002}
    K.~Uchino,
    ``Remarks on the Definition of a Courant Algebroid,''
    Lett.\ Math.\ Phys.\ {\bf 60} (2002) 171
    [arXiv:math/0204010 [math.DG]].

  \bibitem{Vaisman:2004}
    I.~Vaisman,
    ``Transitive Courant algebroids,''
    Int.\ J.\ Math.\ Sci.\ \textbf{2005} (2005) 1737
    [arXiv:math/0407399].

  \bibitem{Freidel:2018tkj}
    L.~Freidel, F.~J.~Rudolph and D.~Svoboda,
    ``A Unique Connection for Born Geometry,''
    Commun. Math. Phys. \textbf{372} (2019) no.1, 119
    doi:10.1007/s00220-019-03379-7
    [arXiv:1806.05992 [hep-th]].

  \bibitem{Marotta:2018myj}
    V.~E.~Marotta and R.~J.~Szabo,
    ``Para-Hermitian Geometry, Dualities and Generalized Flux Backgrounds,''
    Fortsch. Phys. \textbf{67} (2019) no.3, 1800093
    doi:10.1002/prop.201800093
    [arXiv:1810.03953 [hep-th]].

  \bibitem{Freidel:2017yuv}
    L.~Freidel, F.~J.~Rudolph and D.~Svoboda,
    ``Generalised Kinematics for Double Field Theory,''
    JHEP \textbf{11} (2017) 175
    doi:10.1007/JHEP11(2017)175
    [arXiv:1706.07089 [hep-th]].

  \bibitem{Hitchin:2004ut}
    N.~Hitchin,
    ``Generalized Calabi-Yau manifolds,''
    Quart.\ J.\ Math.\  {\bf 54} (2003) 281
      doi:10.1093/qjmath/54.3.281
    [math/0209099 [math-dg]].

  \bibitem{Gualtieri}
    M.~Gualtieri,
    ``Generalized Complex Geometry,''
    Ann. \ of Math. \ {\bf 174} (2011) 75 [arXiv:math/0703298].

  \bibitem{Severa:2001qm}
    P.~Severa and A.~Weinstein,
    ``Poisson geometry with a 3 form background,''
    Prog. Theor. Phys. Suppl. \textbf{144} (2001) 145
    doi:10.1143/PTPS.144.145
    [arXiv:math/0107133 [math.SG]].

  \bibitem{Shelton:2005cf}
    J.~Shelton, W.~Taylor and B.~Wecht,
    ``Nongeometric flux compactifications,''
    JHEP \textbf{10} (2005) 085
    doi:10.1088/1126-6708/2005/10/085
    [arXiv:hep-th/0508133 [hep-th]].

  \bibitem{Blumenhagen:2012pc}
    R.~Blumenhagen, A.~Deser, E.~Plauschinn and F.~Rennecke,
    ``Bianchi Identities for Non-Geometric Fluxes - From Quasi-Poisson Structures to Courant Algebroids,''
    Fortsch. Phys. \textbf{60} (2012) 1217
    doi:10.1002/prop.201200099
    [arXiv:1205.1522 [hep-th]].

  \bibitem{Drinfeld:1986}
    V.~G.~Drinfeld,
    ``Quantum groups,''
    Zap. Nauchn. Sem. LOMI \textbf{155} (1986) 18, J. Soviet Math. \textbf{41}:2 (1988) 898

  \bibitem{Kosmann-Schwarzbach3}
    Y.~ Kosmann-Schwarzbach,
    ``Lie bialgebras, Poisson Lie groups and dressing transformations,''
    Integrability of Nonlinear Systems, Second edition, Lecture Notes in Physics 638, Springer-Verlag, 2004, pp. 107-173.

  \bibitem{Marle}
    C.~M.~Marle,
    ``The Schouten-Nijenhuis bracket and interior products,''
    J. Geom. Phys. \textbf{23} (1997) 350.

  \bibitem{Blumenhagen:2014gva}
    R.~Blumenhagen, F.~Hassler and D.~L\"ust,
    ``Double Field Theory on Group Manifolds,''
    JHEP \textbf{02} (2015) 001
    doi:10.1007/JHEP02(2015)001
    [arXiv:1410.6374 [hep-th]].

  \bibitem{Marotta:2019eqc}
    V.~E.~Marotta and R.~J.~Szabo,
    ``Born Sigma-Models for Para-Hermitian Manifolds and Generalized T-Duality,''
    [arXiv:1910.09997 [hep-th]].

  \bibitem{Ikeda:2020lxz}
    N.~Ikeda and S.~Sasaki,
    ``Global Aspects of Doubled Geometry and Pre-rackoid,''
    [arXiv:2006.08158 [math-ph]].

\end{thebibliography}
\end{document}